\documentclass[letterpaper,titlepage,11pt]{article}
\usepackage{amssymb,amsmath,cite}
\usepackage{times,nicefrac,hhline}

\numberwithin{equation}{section}

\newcommand{\oao}[2]{{#1\atopwithdelims[]#2}}
\newcommand{\oaop}[2]{{#1\atopwithdelims()#2}}
\def\zi{\mathbb{Z}}
\def\er{\mathbb{R}}
\def\ci{\mathbb{C}}

\def\d{{\partial}}
\def\db{\bar{\partial}}
\def\di{\text{d}}

\def\su{SU(2)}
\def\slr{SL(2,\mathbb{R})}
\def\slc{SL(2,\mathbb{R})/U(1)}

\def\p{\partial}
\def\pb{\bar{\partial}}
\def\Id{{\rm 1\kern-.28em I}}

\def\ds{\displaystyle}
\long\def\symbolfootnote[#1]#2{\begingroup%
\def\thefootnote{\fnsymbol{footnote}}\footnote[#1]{#2}\endgroup}

\begin{document}

\begin{titlepage}

\rightline{\vbox{\small\hbox{\tt CPHT-RR087.1108}}}
\vskip 2.5cm

\centerline{\LARGE  Double-Scaling Limit of Heterotic Bundles}
\vskip 0.4cm
\centerline{\LARGE and Dynamical Deformation in CFT}

\vskip 1.6cm
\centerline{\bf L.~Carlevaro$^\diamond$, D.~Isra\"el$^\spadesuit$ and
P.M. Petropoulos$^\diamond$\symbolfootnote[2]{Email:
carlevaro@cpht.polytechnique.fr,israel@iap.fr,petropoulos@cpht.polytechnique.fr}}
\vskip 0.5cm
\centerline{\sl $^\diamond$Centre de Physique Th\'eorique, Ecole Polytechnique,
 91128 Palaiseau, France\footnote{Unit\'e mixte de Recherche
7644, CNRS -- \'Ecole Polytechnique}}

\vskip 0.3cm
\centerline{\sl $^\spadesuit$Institut d'Astrophysique de Paris,
98bis Bd Arago, 75014 Paris, France\footnote{Unit\'e mixte de Recherche
7095, CNRS -- Universit\'e Pierre et Marie Curie}
}

\vskip 1.4cm

\centerline{\bf Abstract} \vskip 0.5cm We consider heterotic
string theory on Eguchi--Hanson space, as a local model of a
resolved A$_1$ singularity in a six-dimensional flux compactification, with an
Abelian gauge bundle turned on and non-zero torsion. 
We show that in a suitable double scaling limit, that isolates the physics near the non-vanishing
two-cycle, a worldsheet conformal field theory description can be
found. It contains a heterotic coset whose target space is
conformal to Eguchi--Hanson. Starting from the blow-down limit of
the singularity, it can be viewed as a dynamical deformation of
the near-horizon fivebrane background. We analyze in detail 
the spectrum of the theory in particular examples, as well 
as the important role of worldsheet non-perturbative effects.

\noindent

\vfill

\end{titlepage}

\tableofcontents

\section{Introduction and Summary of the Results}
Supersymmetric compactifications of the heterotic string~\cite{Gross:1985fr} were 
soon recognized as a very successfull approach to string phenomenology. 
An explicit description at the worldsheet 
conformal field theory (\textsc{cft}) level is only possible at some 
very specific points in the moduli space of compactifications, where 
the geometrical interpretation is usually lost. This includes orbifold 
toroidal compactifications~\cite{Dixon:1986jc}, free-fermionic 
constructions~\cite{Antoniadis:1985az, Antoniadis:1986rn} and Gepner 
models~\cite{Gepner:1987qi}. The topological data of these constructions 
can be mapped onto those of smooth geometrical compactifications, which are 
specified by a given gauge bundle over a Calabi--Yau manifold. For these smooth 
compactifications, although the \textsc{cft} description does not exist, 
it is possible to obtain more realistic $\mathcal{N}=1$ standard-model-like 
spectra, see e.g.~\cite{Braun:2005ux}. Relating the orbifold 
and the smooth description beyond topological data is difficult 
for the full compactification (see~\cite{Honecker:2006qz}), it can however be done in the non-compact 
Calabi--Yau limit, i.e. looking at the neighborhood of a resolved 
orbifold singularity~\cite{Nibbelink:2007rd, Nibbelink:2008qf}.  This approach 
is the same, in spirit, as in type \textsc{ii} models of D-branes at 
singularities~\cite{Aldazabal:2000sa}. 

A crucial role is played in all these approaches by the modified 
Bianchi identity for the  field strength  of the Kalb--Ramond two-form. It should 
include a contribution from the Lorentz Chern--Simons three-form coming from the anomaly-cancellation 
mechanism~\cite{Green:1984sg}, that cannot be neglected in a consistent 
low-energy truncation of the heterotic string:
\begin{equation}
\di \mathcal{H} = \alpha ' \big({\rm tr}\, \mathcal{R}(\Omega_-) \wedge \mathcal{R}(\Omega_-)-\mathrm{Tr}_\textsc{v} \, \mathcal{F} \wedge  \mathcal{F}
\big)\, .
\label{bianchid}
\end{equation}
Consistent torsionless compactifications can be achieved with an embedding of
the spin connexion into the gauge connexion. For more general bundles, 
the Bianchi identity~(\ref{bianchid}) is in general not satisfied locally, but only 
at the topological level.\footnote{Note that the existence of local solutions to 
the Bianchi identity has be shown in some examples in~\cite{Fu:2006vj,Becker:2006et}.} 
One considers usually the necessary (but not sufficient) tadpole-like condition  
$p_1 (T)= \text{ch}_2 (V)$   in cohomology, in terms of the first Pontryagin class of the tangent bundle and 
the second Chern character of the gauge bundle respectively.  This does not probe the local structure of 
the compactification manifold, as~(\ref{bianchid}) would generically lead locally to non-trivial 
Kalb--Ramond fluxes, i.e. manifolds with non-zero torsion. 

Compactifications with torsion were explored in the early days 
of the heterotic string~\cite{Strominger:1986uh,Hull:1986kz} and reconsidered 
recently in the general framework of flux compactifications~\cite{Dasgupta:1999ss,Giddings:2001yu}, 
hoping that they could stabilize the string moduli. Works on heterotic 
flux compactifications 
include~\cite{Curio:2001ae,LopesCardoso:2003af,Becker:2003yv,Becker:2003sh,Manousselis:2005xa,Kimura:2006af}. 
Their analysis is quite involved, as generically the compactification manifold is not even conformally K\"ahler, 
and its topological properties differ from the torsionless manifold. Crucially, 
no smooth compactifications with $\mathcal H \neq 0$ but $\di \mathcal H = 0$ can 
be found~\cite{Ivanov:2000fg}. Therefore generically heterotic compactifications 
should include magnetic sources for the Kalb--Ramond form.  
Such supergravity backgrounds will have also in general a varying dilaton 
over the compactification manifold.

In view of this complexity, it would be useful to be able to describe more quantitatively, 
at least locally, such flux compactifications. In type \textsc{iib} flux 
compactifications~\cite{Giddings:2001yu}, an important role is devoted to throat-like 
regions of the compactification manifold, which are obtained by adding regular and 
fractional D3-brane flux on a deformed conifold-like singularity. Taking into 
account locally the flux backreaction, one obtains approximately a non-compact ten-dimensional 
supergravity solution, whose most celebrated example is the Klebanov--Strassler 
background~\cite{Klebanov:2000hb}. In the 
full model this throat is glued in the \textsc{uv} region to the compactification manifold, in 
a way that is not quantitatively described. However in a suitable decoupling limit one should be 
able to isolate the physics of the throat, barring the fact that the supergravity 
solution may not be weakly curved (as the flux numbers are bounded).

One of the goals of this work is to consider analogous regimes of heterotic flux compactifications, for which part 
of a  compactification manifold  can be blown-up to a full non-compact background, decoupled from the bulk. Such phenomena 
can happen when the gauge bundle degenerates to 'point-like instantons', either at regular points or at singular points, 
leading to strong coupling singularities. These singularities signal the presence of non-perturbative gauge 
groups~\cite{Witten:1995gx}, extra massless tensor multiplets~\cite{Ganor:1996mu}, or both~\cite{Aspinwall:1996vc}. The physics 
in the neighborhood of these singularities  plays an important role in unveiling the full non-perturbative dynamics of 
heterotic compactifications; we expect it to be described by some non-compact and torsional 'near-core' geometry of 
the gauge instantons that become small at the singularity.

In this paper we concentrate on heterotic compactifications to six-dimensions~\cite{Berkooz:1996iz}, that would 
be K3 Calabi--Yau two-folds if the torsion were not present. In this case, one considers
deformed orbifold singularities of the type $\mathbb{C}^2/\Gamma$, which are locally 
described by \textsc{ale} spaces, i.e. self-dual Gibbons--Hawking gravitational instantons~\cite{Gibbons:1979zt}.  
The simpler example is the deformation of a $\mathbb{C}^2/\mathbb{Z}_2$ singularity, known as the 
Eguchi--Hanson instanton.\cite{Eguchi:1978xp}.\footnote{Without fluxes, one obtains a consistent heterotic background 
by embedding the spin connexion into an $SU(2)$ subgroup of the 
gauge group~\cite{Bianchi:1994gi}. The worldsheet \textsc{cft}, though tightly constrained by the 
symmetries, is not solvable.}  We consider $Spin(32)/\mathbb{Z}_2$ heterotic strings on 
this Eguchi--Hanson manifold, with a non-trivial Abelian gauge bundle and non-trivial torsion, tied 
together by the Bianchi identity~(\ref{bianchid}). Unlike the local model of torsion-free 
compactifications of~\cite{Nibbelink:2007rd,Nibbelink:2008qf}, 
we relax the global tadpole constraint $p_1 (T)= \text{ch}_2 (V)$, allowing non-trivial Kalb--Ramond magnetic sources  
even in cohomology (keeping the tadpole-free models as limiting cases).  
Supersymmetric solutions of the heterotic supergravity are found for self- or anti-self-dual gauge field 
backgrounds, together with a fivebrane-like ansatz for the metric, dilaton and Kalb--Ramond field, 
whose transverse space is the Eguchi--Hanson space rather than $\mathbb{R}^4$. It preserves eight supercharges. 
A purely Abelian background of this  sort was considered in~\cite{Cvetic:2000mh} in the limit of large charges,
i.e. neglecting the Lorentz Chern--Simons contribution  
to the Bianchi identity~(\ref{bianchid}).   We will see below that the latter can be satisfied only globally, i.e. 
integrated over the whole non-compact manifold, indicating that this solution is expected to receive 
$\alpha'$ higher-order corrections.

These asymptotically Eguchi--Hanson solutions are similar in spirit 
to asymptotically flat $p$-brane backgrounds. For the purpose of describing locally a compactification space one should isolate 
the throat-like region of this background. Taking the blow-down limit of Eguchi--Hanson, where the bolt shrinks and the
singularity becomes manifest, one finds the near-horizon fivebrane solution, the so-called \textsc{chs} background~\cite{Strominger:1990et}, 
with an extra $\mathbb{Z}_2$ orbifold of the space transverse to the fivebranes, together with a non-trivial monodromy 
of the gauge fields, i.e. a point-like instanton sitting on the singularity. The theory reduces to 
an interacting 'Little String Theory' in six dimensions, of the sort discussed in~\cite{Blum:1997mm}. 
In this regime the smoothness of the background is lost and a strong-coupling singularity emerges.\footnote{Note 
that the \textsc{chs} solution is obtained here in the singular limit of 
Abelian instantons on \textsc{ale}, rather than that of $SU(2)$ instantons on $\mathbb{R}^4$ as 
in~\cite{Strominger:1990et}.}

We exhibit here a novel universal \emph{double-scaling limit} of the line bundle, which allows to 
take a 'near-horizon' limit while keeping the singularity resolved, i.e. its two-cycle non-vanishing. In the same 
spirit as the double-scaling limit of type \textsc{ii} \textsc{ns}5-branes~\cite{Giveon:1999px}, it isolates the non-trivial 
dynamics near the two-cycle. At the same time the string coupling is still \emph{finite everywhere}; 
the tension of the fivebranes wrapping around the two-cycle is held fixed. Geometrically the supergravity solution is 
given as a conformally Eguchi--Hanson manifold, with non-trivial magnetic gauge fluxes through the two-cycles and 
non-trivial torsion. Asymptotically the dilaton is linear, decoupling effectively this throat region from the 
asymptotically (Ricci) flat region. It is therefore a good analogue of a Klebanov--Strassler throat for heterotic 
six-dimensional compactifications. An interesting case corresponds to 
the models whose tadpole condition is satisfied globally for a purely Abelian bundle 
without introducting a fivebrane charge, discussed in~\cite{Nibbelink:2007rd}. We find 
nevertheless in the double-scaling limit of the two-cycle a smooth fivebrane-like solitonic object, corresponding to  
the local geometry of these  seemingly torsion-less models near the bolt of Eguchi--Hanson, becoming 
a singular symmetric fivebrane background in the blow-down limit.

At this stage, one may wonder whether an exact conformal field theory 
exists, that reproduces the heterotic solution at hand (see also~\cite{Nakayama:2008fi}). The
conventional fivebrane background, i.e. the blow-down limit of the gauge bundle, 
is reproduced in the near-horizon limit by (an orbifold of) the $SU(2)$
Wess--Zumino--Witten model (\textsc{wzw}) together with a linear dilaton. 
One may try to take this limit as a starting point for deriving the \textsc{cft} underlying
the resolved background. Asymmetric current--current deformations of the $SU(2)$ 
sigma-model~\cite{Kiritsis:1994fd, Kiritsis:1994ta, Israel:2004vv}
do indeed generate Abelian gauge fields and allow for continuously
squashing the transverse $S^3$ up to the critical limit of $S^2
\times \mathbb{R}$. In the Eguchi--Hanson instanton, the $S^2$ is
present at the bolt, whereas the $S^3$ appears in the asymptotic
region (up to the $\mathbb{Z}_2$). Therefore, in order to describe
the whole Eguchi--Hanson space, we need to promote the
deformation parameter to a function of  the radial transverse coordinate that
connects the bolt with the conical infinity. This is possible
provided the dilaton background is modified accordingly. It is
remarkable that this chain of modifications of the $SU(2)$ \textsc{wzw}
model -- marginal deformation and dynamical promotion -- is
precisely what is needed to reproduce our Eguchi--Hanson plus
Abelian gauge bundle heterotic fivebrane solution in the
double-scaling limit.

Even more remarkable is that the above 
dynamical promotion of the  marginal deformation can actually be identified
with a simple coset conformal field theory. It requires an auxiliary
$SL(2,\mathbb{R})$ \textsc{wzw} model, replacing the linear dilaton 
of the \textsc{chs} model, and is defined as an asymmetric gauged 
\textsc{wzw} model  $[SU(2)_k\times SL(2,\mathbb{R})_k\times (SO(2)_{1,\textsc{r}})^n]/[U(1)_{\textsc{l}}\times U(1)_{\textsc{r}}]$, 
where the factor $(SO(2)_{1,\textsc{r}})^n$ is taken in the free-fermionic current 
algebra of the gauge sector. \footnote{In~\cite{Prezas:2008ua}, an NS5-brane solution in type \textsc{ii} 
transverse to Eguchi--Hanson (with no gauge bundle) was considered, at first order in conformal perturbation theory. 
There the linear dilaton singularity of the fivebranes is not smoothed out.}

Subtleties arise in the definition of this coset, because of the
heterotic nature of the construction. In particular, the freedom
for the amount of superconformal symmetry on the anti-holomorphic
bosonic side of the worldsheet theory allows for two different starting points: $\mathcal{N}=(1,0)$
or $\mathcal{N}=(1,1)$. The second case is merely a particular case of the first one, 
with an Abelian bundle whose first charge is fixed to one unit. This example is nevertheless interesting 
in its own right, as it can be embedded in type \textsc{ii} superstrings in a Kaluza--Klein fashion. 
It gives a coset \textsc{cft} with enhanced $\mathcal{N}=(4,1)$ superconformal symmetry~-- and even 
$\mathcal{N}=(4,2)$ in some cases~-- rather than $\mathcal{N}=(4,0)$ for the generic Abelian bundle. 
In both cases the Bianchi identity~(\ref{bianchid}) is expected to receive corrections beyond the large-charge solution.  
One may be surprised that a sigma-model corresponding 
to a coset \textsc{cft} with extended supersymmetry does receive $\alpha '$ corrections 
to the background fields. We expect that the usual arguments of~\cite{Tseytlin:1992ri,Bars:1993zf} may fail because of the 
asymmetric nature of the coset; the two gauged $U(1)$ factors are even {\it null}. 
We leave the important computation of these corrections for future work.  

Another subtlety concerns the role of the $\mathbb{Z}_2$, which is
necessary, in the field-theory limit, to remove the conical
singularity in the Eguchi--Hanson instanton. Perturbatively (i.e. in conformal perturbation theory),  
nothing like that is required on the worldsheet. However, the $SL(2,\mathbb{R})/U(1)$ coset model is known to
receive worldsheet non-perturbative corrections \cite{fzz,Kazakov:2000pm, Hori:2001ax}. A Sine-Liouville 
type potential is dynamically generated, therefore the non-perturbative consistency of the
\textsc{cft} requires the corresponding operator to be in the physical
spectrum with the appropriate ghost number. In the model that we 
study, this is only possible if the $\mathbb{Z}_2$ orbifold of the $SU(2)$ 
\textsc{wzw} model is present. 

In the supergravity approach to heterotic gauge bundles, an important 
stability constraint (called K-theory constraint in the dual type \textsc{i} language) 
arises~\cite{Witten:1985mj,Freed:1986zx,Blumenhagen:2005ga}. 
It forces the first Chern class of the gauge bundle to be in the 
second even integral cohomology group, in order for the gauge bundle to admit spinors. 
Remarkably, this condition on the Abelian gauge bundle is found in the conformal field theory 
to be related to the \textsc{gso} and orbifold parity of the Liouville potential. 
Only if this condition is satisfied does the Liouville potential 
belong to the physical spectrum. 

The formalism at hand allows to take advantage of the
exact \textsc{cft} description of the Eguchi--Hanson supergravity solution,
in order to determine the full heterotic spectrum and compute the
partition function. This involves both discrete and continuous
representations of the coset $SL(2,\mathbb{R})/U(1)$. The
continuous correspond to asymptotic states concentrated away from
the bolt, whereas discrete states are, as usual, localized in the
vicinity of the resolved singularity. The latter originate from
both untwisted and twisted sectors and are responsible for the
partial breaking of gauge invariance. We study the spectrum 
of massless localized hypermultiplets in two classes of examples. Masses of $U(1)$ gauge fields, 
that arise through the Green--Schwarz mechanism, can also be computed explicitly. 
It would also not be difficult to compute the correlation functions 
for the hypermultiplets localized at the singularity to all orders in $\alpha'$.

This paper is organized as follows. The heterotic supergravity solutions of interest are discussed in sec.~\ref{secsugra}, as well as
their double scaling limit. The sigma-model approach to dynamical deformation is given in sec.~\ref{secsigma}. Section~\ref{seccft} is devoted to the worldsheet conformal field theory description of the backgrounds of interest as gauged \textsc{wzw} models and their worldsheet theories are analyzed. To conclude, we give the explicit massless spectra and gauge symmetries of the vector-like tadpole-free models, raising a puzzle while compairing with previous studies. 
Some details of the supergravity computations are given in app.~\ref{appsigma}. Useful material on
superconformal characters and their identities is gathered in app.~\ref{appchar}.

\section{Heterotic Gauge Bundles over Eguchi--Hanson}
\label{secsugra}

In this section we consider Abelian gauge bundles over Eguchi--Hanson space in heterotic supergravity. 
The solution at lowest order in $\alpha'$ 
can be found explicitly~\cite{Cvetic:2000mh}, and will be analyzed in detail 
in the following. We shall then define the particular double  scaling limit of this solution that will 
eventually be obtained as an exact worldsheet conformal field  theory. We will end this section by 
discussing some limiting cases in which the Bianchi identity can be solved exactly, 
and when one must forgo satisfying it locally, we will present the relevant tadpole conditions for purely 
Abelian bundle in the presence of non-trivial magnetic flux.

\paragraph{Heterotic supergravity}

We start by recalling some facts about heterotic low-energy supergravity, in 
order to fix conventions. The bosonic part of the space-time action reads:\footnote{We are working with anti-hermitian gauge fields, hence the plus sign in front of the corresponding kinetic term. Otherwise, we follow the conventions of~\cite{Duff:1994an}.}
\begin{equation}\label{LagHet}
S = \frac{1}{\alpha'^4}\int \di^{10} x \, \sqrt{-G}\,\mathrm{e}^{-2\Phi}
\left[R + 4|\partial \Phi|^2 -\frac{1}{12} \, |\mathcal{H}_{[3]}|^2 + \frac{\alpha'}{8}
{\rm Tr}_{\text{V}}\, |\mathcal{F}_{[2]}|^2\right]\,,
\end{equation}
the gauge field $\mathcal{A}_{[1]}$ taking values in the vector representation of the gauge group. As usual, 
one can, in place, re-express the gauge sector in the adjoint representation. For $SO(n)$ 
in particular, the identification is ${\rm Tr}_{\textsc{v}}=\frac{1}{n-2} {\rm Tr}_{\textsc{a}}$.\footnote{This is taken 
as a uniform normalisation for both $SO(32)$ and $E_8\times E_8$ case, which is particularly useful for the latter gauge 
group, which lacks a vector representation.} In the following, we will focus on the $SO(32)$ case, i.e. $Spin(32)/\mathbb{Z}_2$ 
heterotic strings, even though our analysis applies to the $E_8\times E_8$ heterotic theory as well.

Invariance of the heterotic supergravity action --~whose bosonic part is given by eq.~(\ref{LagHet})~--
under local supersymmetry transformations plus the implementation of the Green-Schwarz mechanism 
of anomaly cancellation~\cite{Green:1984sg} dictates a modification of the three-form field strength for the Kalb-Ramond two-form. 
This is achieved by the addition of a Yang--Mills and of a (generalized) Lorentz Chern-Simons three-form:
\begin{equation}
\mathcal{H}_{[3]}= \di \mathcal{B}_{[2]}-\alpha' \big(\omega_{[3]}^{\textsc{ym}}(\mathcal{A}) - \omega_{[3]}^{L}(\Omega_-)\big) \, ,
\end{equation}
with
\begin{subequations}
\begin{align}
\omega_{[3]}^{\textsc{ym}}(\mathcal{A}) &= \text{Tr}_{\textsc{v}}\left[ \mathcal{A}\wedge \di\mathcal{A} + \tfrac23 \,\mathcal{A}\wedge\mathcal{A}\wedge\mathcal{A}\right] \, , \\
\omega_{[3]}^{L}(\Omega_-)&=\text{tr}\left[ \Omega_-\wedge \di\Omega_- + \tfrac23 \,\Omega_-\wedge\Omega_- \wedge\Omega_-\right]
\, .
\end{align}
\end{subequations}
The generalized spin connection one-form includes a torsion term generated by the existence of a non-trivial three-form flux:
\begin{equation}\label{genO}
\Omega^{\phantom{\pm}A}_{\pm\,\phantom{A}B} =  \omega^{A}_{\phantom{A}B} \pm \tfrac12 \mathcal{H}^{A}_{\phantom{A}B}\,.
\end{equation}
The modified three-form field strength then satisfies by construction the generalized Bianchi identity:
\begin{equation}
\di \mathcal{H}_{[3]} = 8\alpha'\pi^2 \Big[ \text{ch}_2\big(\mathcal{F}\big) -
 p_1\big(\mathcal{R}(\Omega_{-})\big) \Big]\,,
\label{bianchi}
\end{equation}
in terms of the first Pontryagin class of the tangent bundle and the second Chern character of the gauge bundle. The non-Riemann curvature two-form
 is given in terms of~(\ref{genO}) as 
\begin{equation}
\mathcal{R}(\Omega_{\pm})^{A}_{\phantom{A}B}~=~\di \Omega^{\phantom{\pm}A}_{\pm\,\phantom{A}B} + \Omega^{\phantom{\pm}A}_{\pm\,\phantom{A}C}~\wedge~\Omega^{\phantom{\pm}C}_{\pm\,\phantom{A}B}\, .
\end{equation}

In the absence of fermionic background, a supersymmetric solution to the bosonic equations of motion~(\ref{HETeom}) has to satisfy the 
vanishing of supersymmetry variations for the gaugino, dilaton and gravitino:
\begin{subequations}\label{susy-eq}
\begin{align}
\delta \chi &=  \mathcal{F}_{MN} \,\Gamma^{MN} \epsilon = 0\,,\label{gaugino}\\
\delta \lambda &= -\tfrac{1}{2\sqrt{2}}\,\big(\partial_M \Phi \,\Gamma^M + \tfrac{1}{12}  \mathcal{H}_{MNP} \,\Gamma^{MNP}\big)\,\epsilon = 0\,, \label{dilatino}\\
\delta \psi_M & =\big( \partial_M  + \tfrac14 \Omega_{+\phantom{AB}M}^{\phantom{+}AB} \,\Gamma_{AB}\big)\, \epsilon= 0\, , \label{gravitino}
\end{align}
\end{subequations}
where letters $M,N$ and $A,B$ denote the ten-dimensional coordinate and frame indices, respectively.

\subsection{Gauge bundles on Eguchi--Hanson space}

The resolution of an $A_1$ singularity (i.e. $\mathbb{C}^2/ \mathbb{Z}_2$) is known 
as the Eguchi--Hanson (\textsc{eh}) instanton~\cite{Eguchi:1978xp}. 
This (anti-) self-dual solution of Einstein equations in four dimensions is a Ricci flat and K\"ahler manifold. Its metric 
can be written conveniently by introducing $SU(2)$ left-invariant one-forms:
\begin{subequations}\label{su2-1forms}
\begin{align}
\sigma^\textsc{l}_1 &= \sin \psi_{\mathrm{R}}\, {\rm d} \theta - \cos \psi_{\mathrm{R}} \,
\sin \theta \, {\rm d} \psi_{\mathrm{L}}\,\\
\sigma^\textsc{l}_2 &=  -(\cos \psi_{\mathrm{R}}\, {\rm d} \theta +\sin \psi_{\mathrm{R}} \,
\sin \theta \, {\rm d} \psi_{\mathrm{L}})\,\\
\sigma^\textsc{l}_3 & = {\rm d} \psi_{\mathrm{R}} + \cos \theta \,  {\rm d } \psi_{\mathrm{L}}\,.
\end{align}
\end{subequations}
The \textsc{eh} metric corresponds to a two-center \textsc{ale} space, with the two centers  at a fixed distance $a^2/4$. In terms of the 
left-invariant one-forms~(\ref{su2-1forms}), it reads:
\begin{equation}\label{EHmetric}
\di s^2_\textsc{eh} =  \frac{ \di r^2}{1-\tfrac{a^4}{r^4}} +
\frac{r^2}{4} \left(
(\sigma^\textsc{l}_1)^2 + (\sigma^\textsc{l}_2)^2 +
\Big(1-\frac{a^4}{r^4}\Big) (\sigma^\textsc{l}_3 )^2 \right)\,,
\end{equation}
with the restriction $r\in[a,\infty]$. Thus, the \textsc{eh} manifold possesses an $SU(2)_{\textsc{l}}\times U(1)_{\textsc{r}}$ 
group of isometries, the Abelian factor corresponding to translations along the fiber generated by $\partial_{\psi_{\mathrm{R}}}$.

The topology of \textsc{eh} space is $T^*S^2$. At fixed radius $r$, it can be viewed as  a squashed three-sphere --~considering the 
latter as an Hopf fibration~-- whose deformation is controlled by  
\begin{equation}\label{ehfunc}
g(r)= \sqrt{1-\frac{a^4}{r^4}}\,.
\end{equation}
The space (\ref{EHmetric}) is non-singular in the limit $r \to a$, provided the radius of the $S^1$ fiber is well chosen.  
In this limit there remains a  non-vanishing two-sphere:
\begin{equation}\label{homol-2cycle}
\di s_{\Sigma}^2 = \frac{a^2}{4}\big( \di\theta^2 + \sin^2\theta\,\di\psi_{\mathrm{L}}^2\big)\,,
\end{equation}
which determines a homology basis for the two-cycle $\Sigma$ on the \textsc{eh} space.
In order to avoid a conical singularity at this bolt 
one needs to impose the periodicity $\psi_{\mathrm{R}} \sim 
\psi_{\mathrm{R}} + 2\pi$. Consequently, this solution is asymptotically the orbifold
$\er P_3\equiv S^3/\mathbb{Z}_2$.

Since the manifold is K\"ahler, the existence of a two-cycle homology basis~(\ref{homol-2cycle}) suggests the presence of a corresponding dual (anti-) self-dual two-form. This can be verified 
by integrating the first Pontryagin class of \textsc{eh}, which gives Hirzebruch signature\footnote{There is alternative 
algebraic way of determining the Hirzebruch signature of ALE space, relying on their construction as resolution of $\ci^2/\Gamma$ singularities, 
where $\Gamma\subset \su$ is a discrete Kleinian group, which is $\tau=\text{rk}\,\mathfrak{g}(\Gamma)\,.$ In the case of \textsc{eh}, this gives back the result $\tau=1$.} 
\begin{equation}
\tau(\textsc{eh})=\tfrac13\int p_1(\mathcal R_{\textsc{eh}})=\pm 1\,,
\end{equation}
according to the choice of orientation of the volume measure. Thus, the Eguchi--Hanson space admits one self-dual or anti-self-dual two-form, depending on conventions.

With our choice for the $SU(2)$ left-invariant one-forms~(\ref{su2-1forms}) and for an orientation of the \textsc{eh} space determined by $\varepsilon_{r\psi_{\mathrm{L}}\theta\psi_{\mathrm{R}}}=1$, or in other words by the {\it self-dual} K\"ahler form:
\begin{equation}
J=\big(\hat{e}^0\wedge \hat{e}^3+\hat{e}^1\wedge \hat{e}^2\big)\, ,
\end{equation}
(where the hatted vielbein is restricted to the \textsc{eh} space, see conventions in
app.~\ref{appsigma})
the \textsc{eh} two-form is anti-self-dual, and can be written {\it locally} as:
\begin{equation}
\omega_{[2]} = - \frac{a^2}{4\pi}\di \left(\frac{\sigma^\textsc{l}_3}{r^2} \right)=\frac{a^2}{\pi r^4}\big(\hat{e}^0\wedge \hat{e}^3-\hat{e}^1\wedge \hat{e}^2\big)\,.
\label{harmform}
\end{equation}
Being dual to the two-cycle~(\ref{homol-2cycle}), the two-form~(\ref{harmform}) is then the only 
normalisable representative of the second cohomology class and has intersection number determined by the Cartan matrix of $A_1$:
\begin{equation}
\int_{\Sigma}\omega_{[2]}=1\,,
\qquad \omega_{[2]}\wedge \omega_{[2]} = -\frac{2a^4}{\pi^2 r^8}\,\Omega_{\textsc{eh}}\,,
\qquad \int_{\textsc{eh}}\omega_{[2]}\wedge \omega_{[2]}=-\tfrac{1}{2}\,,
\end{equation}
where $\Omega_{\textsc{eh}}$ (\ref{EHvol}) is the volume form over the \textsc{eh} space. Then from the third relation above, we see that anti-self-duality 
of~(\ref{harmform}) guarantees that $\int \omega_{[2]}\wedge \ast\, \omega_{[2]}$ is positive definite. 

\subsection{Smooth fivebrane-like solution from an Abelian gauge bundle}

Heterotic strings on $\mathbb{R}^{5,1}\, \times\, \textsc{eh}$ can be viewed as a local model of a heterotic K3 compactification 
(which has $\mathcal{N}=1$ supersymmetry in six dimensions)  
in the neighbourhood of an $A_1$ singularity. Following~\cite{Cvetic:2000mh} one 
may add an Abelian gauge bundle proportional to the harmonic two-form~(\ref{harmform}):
\begin{equation}
 \mathcal{F}_{[2]} = -2\pi \sum_{i=1}^{16}\ell_i \,\omega_{[2]} H^i
\label{gaugebund}
\end{equation}
parametrized by charges which can be subsumed in the sixteen-dimensional vector $\vec{\ell}$. In the blow-down limit, $\vec{\ell}$ becomes the 
gauge shift vector encoding the non trivial boundary conditions of $\mathcal{A}$ under the $\zi_2$ orbifold action. 
This gauge field is an  Abelian gauge instanton which takes values in the Cartan subalgebra of $SO(32)$, spanned by the generators $H^i$, $i=1,\ldots,16$.  

As the field strength~(\ref{gaugebund}) is anti-self-dual, 
the vanishing of the supersymmetry variation for the gaugino~(\ref{gaugino}) gives  eight conserved supercharges. In order to determine the chirality of the latter, one decomposes the Majorana-Weyl spinor in ten dimensions 
$\epsilon$~(\ref{susy-eq}) according to the reduced Lorentz group $SO(5,1)\times SO(4)$:
\begin{equation}
16 = (4_+,2_+)\oplus (4_-,2_-)\equiv \epsilon_+\oplus\epsilon_-\,.
\end{equation}
From now on we denote by $\mu,\nu,\ldots$ the $SO(1,5)$ coordinate indices and by $m,n,\ldots$ the $SO(4)$ ones. The Latin letters $a,b,\ldots$ will be used for the $SO(4)$ frame indices.
Then, using the identity $\Gamma^{mn}\,\epsilon_{\pm}=\mp\tfrac12\,\varepsilon^{mn}_{\phantom{mn}rs}\Gamma^{rs}\,\epsilon_{\pm}$, we see immediately that for an anti-self-dual instanton the gaugino variation (\ref{gaugino}) will automatically vanish for $\epsilon_-$.

Finally, the instanton charge of~(\ref{gaugebund}) is fixed by requiring the first Chern class 
to be integral, in order to get a well-defined gauge bundle. One gets 
\begin{equation}
\frac{1}{2\pi}\int_{\Sigma}  \mathcal{F}_{[2]} = -\vec{\ell} \cdot \vec{H}\, ,
\end{equation}
therefore the components $\ell_i$ of the shift vector are either all integer (vectorial shift) or half-integer (spinorial shift), corresponding respectively to gauge bundles with and without vector structure~\cite{Berkooz:1996iz}. 

As we will see later on, 
the fact that the cohomological properties of the \textsc{eh} manifold completely determine the form of the Abelian gauge 
fields tightly constraints tadpole cancellation, thus bearing non-trivial consequences for global solutions to the Bianchi identity~(\ref{bianchi}).

\boldmath
\subsubsection*{Supergravity solution at lowest order in $\alpha'$}
\unboldmath
Because of the non-standard  heterotic Bianchi identity~(\ref{bianchi}), the 
Abelian gauge instanton gives rise to a magnetic source term for the three-form field. 
At lowest order (i.e. in the large charge limit $\vec{\ell}^2 \gg 1$) one can ignore the gravitational 
Chern-Simons term entailed by the Green-Schwarz mechanism. Then eq.~(\ref{bianchi}) yields:
\begin{eqnarray}
\di \mathcal{H}_{[3]} &=& -\alpha' \, \text{Tr}_{\textsc{a}}\,{\mathcal{F}_{[2]}\wedge \mathcal{F}_{[2]}} \notag\\
 &=&-8\alpha'\pi^2\, \vec{\ell}^2 \, \omega_{[2]}\wedge \ast \, \omega_{[2]} \, . 
\label{Hsource}
\end{eqnarray}
Note that the Abelian gauge field is embedded in $SO(32)$ via $U(1)^n\subset SO(2)^n$. In this embedding, the Cartan generator~(\ref{gaugebund}) are anti-Hermitean,
and can be represented by 
\begin{equation}
H^i=\text{diag}\Big( \underbrace{0,\ldots,0}_{2(i-1)},i\sigma^2,\underbrace{0,\ldots,0}_{2(15-i)} \Big)\, ,
\end{equation}
thus accounting for the overall minus sign and the normalisation of the \textsc{rhs} of eq.~(\ref{Hsource}). 

With this in hand, one can write down a supersymmetric solution with non-zero magnetic flux and non-trivial dilaton,  
by resorting to a fivebrane-like ansatz~\cite{Strominger:1990et}, where the fivebrane worldvolume is transverse to the original Eguchi--Hanson space:
\begin{subequations}\label{solH}
\begin{align}
\di s^2 & = \eta_{\mu\nu}\,\di x^\mu \di x^\nu + H(r) \,\di s^2_\textsc{eh} \label{solHa}\,,\\
\mathcal{H}_{[3]} & = g_s^{-2} \mathrm{e}^{2\Phi} \,\ast_{10} \big(\di x^0 \wedge \ldots \wedge \di x^5 \wedge \di H^{-1}\big)\,, \label{solHb}\\
\mathrm{e}^{2 \Phi} & = g_s^{\,2} \, H(r) \label{solHc}\,.
\end{align}
\end{subequations}
This solution verifies the assumption we made in the beginning that the gravity contribution to the Bianchi identity~(\ref{bianchi})  
can be neglected in first approximation. From~(\ref{solH}), we have indeed the relation 
$\mathcal{H}_{mnl} =-2\sqrt{|g|}\varepsilon_{mnl}^{\phantom{mnl}s}\,\partial_s\Phi\sim \mathcal{O}(\alpha')$, so that the gravity 
contribution in (\ref{bianchi}) scales like  $\text{Tr}\,\mathcal{R}_-\wedge\mathcal{R}_-\sim \mathcal{O}(\alpha'^2)$, which is a third order
correction to the Bianchi identity.

The profile of the conformal factor $H$ on Eguchi--Hanson can be determined from the equations of motions and satisfies the 
transverse space Laplace equation on Eguchi--Hanson with a source:
\begin{equation}
\Delta_\textsc{eh} H(r) = -\frac{16\alpha ' a^4 \vec{\ell}^2}{r^8}\,.
\label{laplace}
\end{equation}
A simple $SU(2)_{\textsc{l}}\times U(1)_{\textsc{r}}$ invariant asymptotically flat solution to the above is provided by:
\begin{equation}\label{con-fac}
H(r) = 1+ \frac{2\alpha'\vec{\ell}^2 }{r^2}.
\end{equation}
It is interesting to note that this solution gives in~(\ref{laplace}) an extra delta-function 
fivebrane source localized at $r=0$, even though this point is not part of the manifold (which is covered by $r\geqslant a$).

Last but not least, one has to verify that solution ~(\ref{solH}) preserves the same set of supercharges as the gauge instanton~(\ref{gaugebund}). 
Because of the relations~(\ref{solHb}, \ref{solHc}) linking the three-form to the dilaton, the dilatino variation~(\ref{dilatino}) 
simplifies to $\delta\lambda=-\tfrac{1}{2\sqrt{2}}\partial\Phi\cdot(\Gamma_1+(\ast \Gamma_3))\,\varepsilon_{\pm}$.
By resorting to the identity $\Gamma^{m}\,\epsilon_{\pm}=\pm\tfrac16\,\varepsilon^{m}_{\phantom{m}nrs}\Gamma^{nrs}\,\epsilon_{\pm}$, 
we verify that eq.~(\ref{dilatino}) vanishes for $\epsilon_-$. The story can be repeated 
for the gravitino variation~(\ref{gravitino}). As will be explained in sec.~\ref{sec:BI}, $\Omega_+$ is anti-self-dual, which, 
combined with the duality properties of $\Gamma^{mn}\epsilon_{\pm}$, will make $\delta\lambda$ vanish for $\epsilon_-$ only.

To conclude, we want to stress how tightly correlated are the duality property of the \textsc{eh} space,  hence of the 
associated Abelian gauge instanton, the sign of $\mathcal{H}$ and the necessity of always using different generalized spin 
connections in the Bianchi identity and in the supersymmetry variations. There are only two possible combinations (summarized below) depending on whether we choose the \textsc{eh} space to be self- or anti-self-dual. 
This excludes any arbitrariness in the choice of signs:
\begin{equation}
\begin{array}{cccc}
& \qquad& \textsc{Bianchi}\qquad & \textsc{Susy}\\
\mathcal{F}_{[2]} =\pm \ast_4\mathcal{F}_{[2]}\,,&\qquad
\mathcal{H}_{[3]} =\pm 2\ast_{10}\!\big(\text{Vol}(\er^{1,5})\wedge \di \Phi\big)\,,\qquad & \Omega_{\pm}\,, \qquad&\Omega_{\mp}\,.
\end{array}
\end{equation}
These two alternative backgrounds both preserve 8 chiral/antichiral supercharges, while the other remaining half is associated to the fermionic zero-modes bound to the instanton.

Before tackling the possibility of exactly solving the Bianchi identity~(\ref{bianchi}) locally, we will dwell a bit on two limits of 
the background~(\ref{solH}), which will be instrumental in the following.
 
\subsubsection*{Fivebranes as the blow-down limit}
One can consider the blow-down limit of Eguchi--Hanson space for the solution~(\ref{solH}), defined as 
$a\to 0$ for $g_s$ fixed. eq.~(\ref{solH}) gives then the standard  heterotic symmetric fivebrane solution, 
with  $\mathbb{C}^2/\mathbb{Z}_2$ as transverse space. In this limit the gauge bundle degenerates to a 
non-trivial monodromy for the gauge field around the singularity, that we will determine in sec.~\ref{seccft}. 

This heterotic string background has a strong coupling singularity for $r\to 0$. It is important that this 
blow-down limit does {\it not} give the worldsheet \textsc{cft} for $\mathbb{C}^2/\mathbb{Z}_2$ at the orbifold point. 
In that case, the dilaton is constant as the fivebrane charge vanishes. The orbifold point is obtained 
for the standard embedding of the spin connexion into the gauge connexion that we discuss below.

We recall that the heterotic fivebrane is also known to be obtained in the singular zero-size limit of 
non-Abelian gauge instantons on the transverse $\mathbb{R}^4$~\cite{Strominger:1990et}. We have here 
a similar mechanism for an Abelian instanton on a transverse Eguchi--Hanson. In the present case the instanton size, rather than 
being an independent parameter, is set by the resolution parameter of the two-cycle.

\subsubsection*{Double scaling limit} 
One can define a very interesting alternative limit of this supergravity solution. One would like to isolate 
the dynamics near the two-cycle of the resolved singularity, without going to the blow-down limit, i.e. keeping 
the transverse space to be conformal to a non-singular Eguchi--Hanson space. 
This can be achieved by taking the following {\it double scaling limit}:
\begin{equation}\label{DSL}
g_s \to 0 \quad , \qquad \frac{a}{g_s \sqrt{\alpha'}} \quad \text{fixed}\,.
\end{equation}
In this limit the tension of five-branes wrapped around the two-cycle of Eguchi--Hanson is kept fixed, being 
proportional to $\mathcal{V}_{\Sigma}/g_s^2$. One obtains then an interacting theory whose effective coupling 
constant is set by the double scaling parameter. 

In this limit it is convenient to change from the radial coordinate to $\cosh \rho = (r/a)^2$, that is held fixed 
in the process. Introducing the quantized fivebrane charge:
\begin{equation}\label{N5flux}
 Q_5 = -\frac{1}{4\pi^2 \alpha'} \int_{\er P_3 , \, \infty}\!\!\!\!\!\! \mathcal{H}_{[3]}\,,
\end{equation}
we obtain the metric
\begin{subequations}\label{eqDSL}
\begin{align}
\di s^2 & = \eta_{\mu\nu}\,\di x^\mu \di x_\nu +  \frac{\alpha' Q_5}{2}\Big[ {\rm d}\rho^2 + (\sigma^\textsc{l}_1)^2 +
(\sigma^\textsc{l}_2)^2 + \tanh^2 \rho \, (\sigma^\textsc{l}_3 )^2\Big]\\
\mathcal{H}_{[3]} & = -\frac{\alpha' Q_5}{2} \tanh^2 \rho\,\, \Omega_3 \\
\mathrm{e}^{2 \Phi} & = \frac{2 g_s^2 \,\alpha ' Q_5}{a^2} \frac{1}{\cosh \rho} \label{dilateq}\\
\mathcal{F}_{[2]} & =    \di \left( \frac{\sigma^\textsc{l}_3}{2\cosh \rho}\right)\, \vec{\ell} \cdot \vec{H}\, , 
\end{align}
\label{sugrasol}
\end{subequations}
$\Omega_3$ being the volume form on $S^3$ with orientation determined by $\varepsilon_{\psi_{\mathrm{L}}\theta\psi_{\mathrm{R}}}=1$.
In section~\ref{seccft}, we will be able to recover this heterotic supergravity solution as an exact worldsheet conformal field theory. 

\subsection{Bianchi identity}\label{sec:BI}

On general grounds, higher order $\alpha'$ corrections to a heterotic supergravity background are proportional to the Bianchi identity or 
contractions of thereof, so an exact and consistent such background should satisfy~(\ref{bianchi}), if not locally, at least globally, in cohomology.

Firstly, let us try and figure out how possible it is to solve (\ref{bianchi}) locally. 
As seen previously in eq. (\ref{Hsource}), one cannot rely on some of the Abelian gauge factors to achieve this, 
since their contribution to~(\ref{bianchi}) goes, for the present background, like $1/r^5$ --~as they can be only proportional 
to the harmonic two-form~(\ref{harmform})~--  while the leading term from the curvature contribution scales like $1/r^9$. The only way out is to cancel the 
contribution from curvature with torsion against a non-Abelian gauge fibration, by attempting a standard-like type embedding of the gauge connection into the 
Lorentz spin connection with torsion. This will require coupling non-Abelian gauge fields stemming from $SO(32)$ to the background, while the Abelian 
gauge fields from the fibration  will only contribute to sourcing non trivial magnetic flux.

The conditions for a standard embedding rely on the duality properties of the generalized spin connection with torsion. These can be made overt by decomposing:
\begin{equation}\label{om2}
\Omega^{\phantom{\pm}a}_{\pm\,\phantom{a}b} =  
\hat{\omega}^{\phantom{\textsc{eh}}a}_{\textsc{eh}\phantom{a}b} + \big(\hat{e}_{a} \,\hat{e}_{b}^{\phantom{b}\,\mu}- 
\hat{e}_{b} \,\hat{e}_{a}^{\phantom{b}\,\mu}\big)\,\partial_\mu \Phi \pm \tfrac12 \mathcal{H}^{a}_{\phantom{a}b}\,
\end{equation}
where (hatted) quantities are restricted to the (un)warped \textsc{eh} space (cf.~(\ref{vier}) and~(\ref{omegaEH})), and the torsion one-form is given by:
\begin{equation}
\mathcal{H}_{ij}=-4\alpha' Q_5 \,  g(r)(r^2H(r))^{-3/2} \epsilon_{ijk}\,e^k\,, \quad \text{and} \quad \mathcal{H}_{0i}=0\,,\quad  i,j,k=1,2,3\,. 
\end{equation}
In particular, keeping in mind the anti-self-duality of $\hat\omega_{\textsc{eh}}$ and recasting the \textsc{nsns} flux in terms of hatted indices as $\mathcal{H}_{ab}= -2\,\epsilon_{abcd} \,\hat{e}^{c} 
\hat{e}^{d\mu}\,\partial_\mu \Phi$, we observe that $\Omega_-$ has 
no overall (anti)-self-duality properties, while $\Omega_+$ is anti-self-dual.

By turning on a $SU(2)\times SU(2)\cong SO(4)\subset SO(32)$ gauge field, one can in principle define four different embeddings of the gauge connection into 
the modified Lorentz connection~\cite{Bianchi:1994gi}:
\begin{equation}
\overset{\, (-)}{\mathcal{A}}\!\!\left.^i_{\pm\,[1]}\right. = \tfrac12\,\overset{(-)}{\eta}\!\!\left.^i_{ab}\right.\,\Omega^{\phantom{\pm}ab}_{\pm} \, ,
\end{equation}
via contraction with the (anti-)self-dual t'Hooft symbols $\eta^i_{ab}$, $\bar\eta^i_{ab}$, $i$ being in the adjoint of $\su$.

However, if the four above embeddings are all in principle admissible, only $\mathcal{A}_{-}$ and $\bar{\mathcal{A}}_{-}$ are good potential 
candidates. One immediately sees, for instance, that as $\Omega_+$ is anti-self-dual, $\mathcal{A}_+=0$. Furthermore, even though $\bar{\mathcal{A}}_+$ 
does not vanish and can be embedded into the entire $\Omega_+$ without further effort, the corresponding field strength 
$\bar{\mathcal{F}}_{+}=\tfrac14\,\bar\eta_{ab}\mathcal{R}^{ab}_{\phantom{ab}\mu\nu}(\Omega_+)\,\di x^{\mu}\wedge \di x^{\nu}$ is not
anti-self-dual,\footnote{Let us verify this in the near horizon limit, for simplicity. Since $\Omega_+$ is anti-self-dual, we can perform the 
standard embedding with a single $SU(2)\subset SO(32)$. We obtain $\mathcal{F}_{[2]}=a^4\, \di\left( \tfrac{\sigma_3^{\textsc{l}}}{r^4}\right)\,i\sigma^3$ 
which is in this case Abelian but, obviously, cannot be anti-self-dual, since it is not proportional to $\omega_{[2]}$. In the asymptotically flat case, the 
gauge field-strength becomes truly non-Abelian, but the conclusion about the absence of duality property remains unchanged.} since 
$\mathcal{R}^{ab}_{\phantom{ab}\mu\nu}$ is anti-self-dual in the frame but not in the coordinate indices.\footnote{This was shown in~\cite{Bianchi:1994gi} when $\di \mathcal{H}= 0$, but extends here to a case with non trivial three-form flux.}

This intrinsic relation between the semi-simpleness of the Euclidean Lorentz group $SO(4)$ and the duality properties of $\Omega_{\pm}$ 
brings about a splitting of the gravity term in~(\ref{bianchi}), manifesting a sort of orthogonality  between (anti-)self-dual 
pieces of $\mathcal{R_{\pm}}$ under the first Pontryagin class:
\begin{equation}\label{p1}
p_1\Big(\mathcal{R}(\Omega_{\pm})\Big) = 
p_1\Big(\mathcal{R}(\hat\omega_{\textsc{eh}})\Big) 
\mp p_1\Big(\mathcal{R}(\Omega_{-}-\hat\omega_{\textsc{eh}})\Big)\,.
\end{equation}

We can now determine the gravity correction with torsion to the Bianchi identity for the supergravity background~(\ref{solH}), 
with the asymptotically flat conformal factor $H$~(\ref{con-fac}). In order to make contact later on with the 
near horizon geometry, we slightly generalize the 
conformal factor to  $H(r)=\lambda + \frac{2\alpha'Q_5}{r^2}$ with $\lambda\in[0,1]$. Then, the double-scaling limit (\ref{DSL}) is rephrased as $\lambda\rightarrow 0$, while $\lambda=1$ restores the full asymptotically flat supergravity background. After some manipulations, 
the topological term (\ref{p1}) can be written in terms of the conformal factor $H(r)$ and the 
function $g(r)$ in the \textsc{eh} metric (\ref{ehfunc}) as follows:
\begin{multline}\label{pontryO}
\text{tr}\big(\mathcal{R}(\Omega_{\pm})\wedge \mathcal{R}(\Omega_{\pm})\big)\\=
 \left\{192 \left(\frac{(1-g^2)^2}{r^{4}}\right)\, \pm \,32\,\frac{(H-\lambda)^2}{r^4H^4}
\Big[2(1-g^2)H\big[(1-g^2)H+\lambda g^2\big]-3\lambda^2 g^4 \Big]\right\}\,\Omega_{\textsc{eh}}\, .
\end{multline}

Before turning to the possibility of performing a standard embedding, a few comments are in order on eq.~(\ref{pontryO}).
\begin{itemize}
\item In the asymptotically flat case ($\lambda=1$), the second term on the \textsc{rhs} of eq.~(\ref{pontryO}) integrates to zero, leading to the Hirzebruch signature:
\begin{equation}\label{Hirzpm}
\tau_{\pm}=\frac13\int_{\mathcal{M}}p_1\big(\mathcal{R}(\Omega_{\pm})\big)=\tau(\mathcal{M}_{\textsc{eh}})=-1\,.
\end{equation}
Adding torsion and warping does not change, in this case, the cohomological properties of the generalized metric with respect
to \textsc{eh}.
\item The near horizon limit of expression~(\ref{pontryO}) can be computed by taking $\lambda\rightarrow 0$:
\begin{equation}\label{top1}
\text{tr}\big(\mathcal{R}(\Omega_{\pm})\wedge \mathcal{R}(\Omega_{\pm})\big) |_{\text{n.h.}}= 64\,(3\pm 1)\left(\frac{a^8}{r^{12}}\right)\,\Omega_{\textsc{eh}}.
\end{equation}
In this regime, the contribution from torsion terms is proportional to the pure \textsc{eh} one. 
However, the non-linearity of $\mathcal{F}$ w.r.t. $\mathcal{A}$  does not 
allow to cancel both contribution in the Bianchi identity by a simple rescaling of a common gauge field $\mathcal{A}$.
We also notice that for $\Omega_{-}$, which is, in our conventions, relevant for the Bianchi identity, 
the topological term~(\ref{top1}) decreases by one unit.
\item
The blow-down limit can be considered by taking $a\to 0$, i.e. $g\rightarrow 1$. Then~(\ref{pontryO}) is non-vanishing only in the case where the conformal factor is asymptotically flat:
\begin{equation}
\text{tr}\big(\mathcal{R}(\Omega_{\pm})\wedge \mathcal{R}(\Omega_{\pm})\big)= \mp\left(\frac{384\,(\alpha' Q_5)^2}{r^8 H^4}\right)\,\Omega_{\textsc{eh}}\,
\end{equation}
while, as expected, it vanishes altogether in the near horizon limit for $r>0$, since in this regime
$\Omega_+=0$ and $\Omega_-$ is a flat connection (see app.~\ref{appsigma}).
\end{itemize}
\subsubsection*{Standard embedding}

As it has been argued above, the only embedding of a non-Abelian gauge field into the generalized spin connection which is compatible with supersymmetry is given by $\mathcal{A}_-$. 
The two possible mutually commuting embeddings are:
\begin{eqnarray*}
\mathcal{A}_{[1]\,-}&=&  \tfrac12\,\eta_{ab}^k \,\big( 2\hat{e}^{a} \,\hat{e}^{b\,\mu}\,\partial_\mu \Phi \pm \tfrac12 \mathcal{H}_{ab}\big)\big(\tfrac{i}{2}\sigma_k\big)\,,\\[5pt]
\bar{\mathcal{A}}_{[1]\,-}&=&\tfrac12\,\bar\eta_{ab}^k\,\hat\omega^{\phantom{\textsc{eh}}ab}_{\textsc{eh}}\,\big(\tfrac{i}{2}\bar\sigma_k\big)\,,
\end{eqnarray*}
the t'Hooft symbols realising the decomposition of the Euclidean Lorentz group $SO(4)$ onto the generators of $SU(2)\times SU(2)$:
\begin{equation}
\overset{(-)}{\sigma}\!\!\left.^k \right.= -i \overset{(-)}{\eta}\!\!\left.^k_{IJ}\right.\,\Sigma^{IJ}\,,\qquad 
\overset{(-)}{\eta}\!\!\left.^i_{IJ} \right.= \mp\, 2\delta_{[I}^0\delta_{J]}^i+ \varepsilon^{i}_{\phantom{i}jk}\delta^j_{[J} \delta^k_{K]}\,,
\end{equation}
where $\Sigma_{IJ}$ are the generators of the adjoint of $SO(4)$, satisfying $[\Sigma_{IJ},\Sigma_{KL}]=2\delta_{I[L}\Sigma_{K]J}-2\delta_{J[L}\Sigma_{K]I}$. 
An embedding with both $\su$ factors turned on would clearly cancel against the whole gravity contribution~(\ref{pontryO}).
However, one must still verify that the {\it whole} resulting $SO(4)$ gauge field-strength is anti-self-dual.

Firstly we consider the embedding of the first $\su$ factor into the self-dual part of $\Omega_-$, denoted
$\tilde\Omega=\Omega_- - \hat\omega_{\textsc{eh}}$ for simplicity. The curvature two-form reads in this case:
\begin{subequations}
\begin{align}
\mathcal{R}^{01}(\tilde\Omega)&= \mathcal{R}^{23}(\tilde\Omega)=\frac{2(H-\lambda)}{r^2 H^2}\Big(
\big[H(1-g^2)-\lambda g^2\big]\,\hat{e}^0\wedge\hat{e}^1 + \big[H(1-g^2)+\lambda g^2\big]\,\hat{e}^2\wedge\hat{e}^3
\Big)\,\notag\\
 \mathcal{R}^{02}(\tilde\Omega)&=-\mathcal{R}^{13}(\tilde\Omega)=\frac{2(H-\lambda)}{r^2 H^2}\Big(
\big[H(1-g^2)-\lambda g^2\big]\,\hat{e}^0\wedge\hat{e}^2 - \big[H(1-g^2)+\lambda g^2\big]\,\hat{e}^1\wedge\hat{e}^3
\Big)\,\notag\\
\mathcal{R}^{03}(\tilde\Omega)&=\mathcal{R}^{12}(\tilde\Omega)=\frac{2(H-\lambda)}{r^2 H^2}\Big(
\big[2H(1-g^2)-\lambda g^2\big]\,\hat{e}^0\wedge\hat{e}^3 + \lambda g^2\,\hat{e}^1\wedge\hat{e}^2
\Big)\,\notag
\end{align}
\end{subequations}
which is anti-self-dual for coordinate indices only in the blow-down regime $g\rightarrow 1$. However, in this regime the double scaling limit~(\ref{DSL}) which is relevant for our 
\textsc{cft} description is trivial. Anyhow, since the pure \textsc{eh} contribution to the first Pontryagin class~(\ref{pontryO}) vanishes in the blow-down limit (up to a delta-function localized term at $r=0$) the standard embedding can be performed with a single $\su$, 
in the non trivial case where the conformal factor is taken to be asymptotically flat.
In the blow-down limit, the standard embedding yields:
\begin{equation}\mathcal{A}^i_-=\frac{2(H-1)}{rH}\,\hat{e}^i\,,
\end{equation}
leading to an anti-self-dual field-strength:
\begin{equation}
\mathcal{F}_{[2]\,-}^i= -\frac{4(H-1)}{r^2H^2}\,\left(\hat{e}^0\wedge \hat{e}^i -\tfrac12\,\varepsilon^i_{\phantom{i}jk}\,\hat{e}^j\wedge \hat{e}^k\right)\,.
\end{equation}
This cancels the second contribution in the topological term~(\ref{pontryO}), while the pure \textsc{eh} contribution vanishes in the blow-down limit. 
We obtain  a standard heterotic fivebranes background, whose transverse space is the quotient $\ci^2/\zi_2$.

Secondly the embedding of the second $\su$ factor via $\bar{\mathcal{A}}_-$ corresponds to the pure \textsc{eh} case, which is consistent only with a constant dilaton and no torsion, and subsequently no flux.\footnote{ 
This setup was studied in detail in~\cite{Bianchi:1994gi}; some information is given in app.~\ref{appsigma}.}
Suffice to have a look at the \textsc{eh} curvature two-form~(\ref{R2EH}) to convince oneself that $\bar{\mathcal{F}}^i=
\tfrac12\bar\eta_{ab}^i\,\mathcal{R}^{ab}(\hat\omega_{\textsc{eh}})$ is anti-self-dual.

To summarize, we observe that we can only turn on one $\su$ gauge field at a time and be in keep with supersymmetry. This is ascribable to the reduction of the 
isometry group of the four-dimensional transverse space from $SO(4)$ to $SU(2)_{\textsc{L}}\times U(1)_{\text{R}}$. Then, there are only two distinct regimes 
in which the Bianchi identity can be solved locally for the background~(\ref{solH}) and an $\su$ gauge field coupled to it. These regimes exclude each other:
\begin{itemize}
\item The pure \textsc{eh} case, with constant dilaton and without any Abelian gauge instanton.
\item The blow-down limit of heterotic fivebranes transverse to \textsc{eh}, with asymptotically flat conformal factor. The Abelian gauge field collapses to a point-like instanton.
\end{itemize}

The corresponding sigma-model description for these two cases has enhanced $\mathcal{N}=(4,4)$ superconformal symmetry, which accounts for the absence of perturbative corrections to 
the $\beta$-functions. However, their \textsc{cft} description is far from obvious.\footnote{In~\cite{Anselmi:1993sm}, a partial identification of \textsc{ale} instanton backrounds with 
deformations of solvable $\ci^2/\Gamma$ \textsc{cft}'s  has been established, however in a particular singular limit.}  
The full \textsc{cft} for the resolved $\ci^2/\zi_2$ orbifold without fivebranes has still to be found. 
As for the second case, it is clearly out of the range of validity of the \textsc{cft} description carried out in the present work, 
except in the near-horizon limit~\cite{Callan:1991dj}. In conclusion,
the exact \textsc{cft} we will work out in sec.~\ref{seccft} applies to neither of these two cases, 
however exact they can be from a supergravity perspective.

The outcome of the previous analysis is that in the double scaling limit~(\ref{DSL}), where a \textsc{cft} description is possible as we shall see, 
we expect the supergravity background~(\ref{eqDSL})  to receive perturbative $\alpha'$ corrections to the background fields, with an expansion parameter $1/Q_5 \sim 1/\vec{\ell}^2$ (in a related model, such $\alpha'$ corrections have been computed as a series expansion~\cite{Fu:2008ga}).  Moreover, as for a generic 
shift vector $\vec{\ell}$ no non-Abelian sub-group of $SO(32)$ is preserved by the bundle, 
we do not expect any additional non-Abelian gauge fields to be added to the background as a correction. 
Finally, because of the presence of a nontrivial resolution two-cycle, worldsheet instanton corrections 
are likely as well. They will actually be found in the \textsc{cft} description. 

\subsubsection*{Abelian fibration: tadpole cancellation condition}

Since the heterotic supergravity background given by eqs.~(\ref{eqDSL}) does not possess a consistent standard embedding solution to make it exact in a regime where both fivebranes are present 
and the orbifold is resolved, we can relax the locality condition for the Bianchi identity, and
require it only to be globally fullfiled, and therefore also in cohomology. The integrated version of~(\ref{bianchi}) yields
\begin{equation}
\frac{1}{4\pi^2 \alpha'} \int_{\er P_3 , \, \infty}\!\!\!\!\!\! \mathcal{H}_{[3]}
= 2 \Big[\int_{\mathcal{M}} \text{ch}_2\big(\mathcal{F}\big)-3\,\tau\big(\mathcal{R}(\Omega_{-})\big) \Big]
\end{equation}
leading to the following tadpole cancellation condition, in the asymptotically flat case:
\begin{equation}
Q_5 = \vec{\ell}^2 -6\, . 
\end{equation}
In the absence of flux, we recover the same condition as in~\cite{Nibbelink:2007rd}, which is reminiscent of the consistency relation for fractional branes~\cite{Conrad:2000tk}.  
In the double scaling limit, this condition becomes:
\begin{equation}\label{tad}
Q_5|_{\text{n.h.}} = \vec{\ell}^2 -4 = Q_5 + 2\,.
\end{equation}
The tadpole condition~(\ref{tad}) will be relevant for the  worldsheet \textsc{cft} description we will give of the background~(\ref{eqDSL}), in sec.~\ref{seccft}.

\section{Sigma-Model Approach: Dynamical Deformations}
\label{secsigma}

In this section we will uncover the worldsheet sigma-model structure of the supergravity background~(\ref{sugrasol}), 
as a first step towards determining the underlying exact worldsheet conformal field theory. We will find that, 
starting from the blow-down limit of the gauge bundle over Eguchi--Hanson, one can obtain the 
resolved singularity with an Abelian gauge bundle by resorting to the method of dynamical promotion of a marginal deformation, called hereafter in short \emph{dynamical deformation}. 
This technique consists in  giving a field dependence to the parameter of an exactly marginal deformation in a worldsheet \textsc{cft}.\footnote{
In the literature, this issue
has been addressed only for the time-dependent case, aiming at
generating cosmological-like backgrounds
\cite{Kiritsis:1994fd}.}

\subsection{Marginal deformations and dynamical promotion}

Let us consider a string sigma-model admitting some target space isometries realized as 
chiral currents $\mathcal{J}^a (z)$ and $\tilde{\mathcal{J}}^b (\bar z)$, which generate (possibly
non-Abelian) left-moving and right-moving affine algebras (in the case of an homogeneous space, one gets a Wess--Zumino--Witten 
(\textsc{wzw}) model). Marginal deformations of the original conformal field theory are realised using
left-right combinations of these currents:
\begin{equation}
\delta S = \frac{1}{2\pi \alpha' } M_{AB} \int \di^2 z \, \mathcal{J}^A (z) \tilde{\mathcal{J}}^B (\bar z)\, .
\label{marg}
\end{equation}
The matrix $M_{AB}$ has components only along the Cartan of the left- and right-algebra (which 
guarantees the deformation to be exactly marginal~\cite{Chaudhuri:1988qb}) and defines  
a new conformal field theory for arbitrary entries thereof. The idea behind {\it dynamical deformation} 
is to replace this constant matrix by a field-dependent one. In the simplest case, one adds to the sigma-model 
a spectator bosonic field $\varrho(z,\bar z)$ and the worldsheet action is modified as
\begin{equation}
\delta S = \frac{1}{2\pi \alpha'}\int \di^2 z\, \left(\p \varrho \,\pb \varrho +  \mu(\varrho) M_{AB}  \mathcal{J}^A  \tilde{\mathcal{J}}^B + \Phi (\varrho)\,R_{(2)}\right) \, , 
\end{equation}
where, anticipating a little bit our results, we included a $\varrho$-dependent dilaton  $\Phi (\varrho)$, which 
is in general necessary to preserve conformal invariance, since a new coordinate $\varrho$ now enters non-trivially in the geometry. 
A solution to the beta-function equations for $\mu(\varrho)$ and $\Phi (\varrho)$ gives a   
{\it dynamical promotion} of the marginal deformation~(\ref{marg}). 

A more involved issue is whether the new sigma-model obtained after implementing the dynamical deformation procedure 
possesses a description in terms of a manifestly exact \textsc{cft}. A known example  is the promotion
of a symmetric deformation of the $SU(2)_k$ \textsc{wzw} model:
\begin{equation}
S = S_{SU(2)_k}+\frac{1}{2\pi \alpha'}\int \di^2 z\,\left(\p \varrho \,\pb \varrho +  \mu(\varrho) J^3 \bar{J}^3 + \Phi  (\varrho)\,R_{(2)}\right) \,.
\end{equation}
A conformal field theory description of the latter is given as the gauged \textsc{wzw} model $[SU(2)_k\times \slr_k] /(U(1)_{\textsc{l}}\times U(1)_{\textsc{r}}$~\cite{Israel:2004ir}.\footnote{Or 
after T-duality by the orbifold $[\slc \times SU(2)/U(1)]/\mathbb{Z}_k$, see~\cite{Sfetsos:1998xd,Giveon:1999px}.}
By embedding this 
\textsc{cft} in a full string-theory setup, one recognizes the background of  $k$ NS5-branes  spread on a topologically 
trivial circle~\cite{Sfetsos:1998xd}. Hence, the configuration of NS5-branes on a circle can  be beautifully obtained by dynamically promoting
a marginal deformation of the Callan--Strominger--Harvey (\textsc{chs}) background.\footnote{There is an alternative viewpoint to the dynamical 
promotion of an $SU(2)$ marginal deformation. In the framework of supersymmetric \textsc{wzw} or gauged  \textsc{wzw} models combined with a 
linear-dilaton background, a large spectrum of marginal deformations is available. These deformations have been studied in detail 
in~\cite{Fotopoulos:2007rm, Prezas:2008ua}, where it was shown how they are related to the deformation of the distribution of the 
branes. Hence, it is possible to move to the circle distribution, from the system of  branes concentrated on a point, by a purely 
marginal deformation involving also the dilaton vertex operators.}  The coordinate $\varrho$ requested for the 
dynamical promotion also finds a natural interpretation as the linear dilaton support of the undeformed \textsc{chs} background.

In the following sections, we will work out a new example, where both the dynamical promotion and 
the exact \textsc{cft} description can be obtained.  Our  starting point will be the 
blow-down limit of the supergravity solution~(\ref{solH}) in the near-horizon 
heterotic fivebranes solution, as its conformal field 
theory description is well-known. Setting aside the $\mathbb{Z}_2$ orbifold (that 
plays no role in the present sigma-model analysis), it corresponds to the  
so-called Callan--Strominger--Harvey (\textsc{chs}) background~\cite{Kounnas:1990ud,Callan:1991dj}:
\begin{equation}\label{chs-het}
\mathbb{R}^{5,1} \times \mathbb{R}_Q \times SU(2)_{k}\,.
\end{equation}
The linear dilaton has a background charge ${\mathcal Q}=\sqrt{2/\alpha' k}$  determined by 
the level $k$ of the current algebra. When embedded in a full-fledged heterotic background, 
this \textsc{cft} describes the near-horizon geometry of a configuration of parallel and coincident 
heterotic fivebranes. In the following, we will show that by starting from the \textsc{chs} background~(\ref{chs-het}), 
deforming it by a marginal current--current deformation that turns on a gauge field in target space 
and promoting the latter dynamically, we recover the supergravity solution~(\ref{sugrasol}). 

\subsection{Asymmetric marginal deformation of the heterotic fivebrane}
The starting point is the \textsc{wzw} action on an Euclidean surface $\mathcal{S}$, for the group-valued field $g \in SU(2)$:
\begin{equation}
S_G = \frac{k}{8\pi}\int_\mathcal{S}\!\! \di^2 z\, {\rm Tr}(\partial g \bar{\partial} g^{-1}) 
+ \frac{ik}{24\pi}\int_\mathcal{B} {\rm Tr} (g^{-1} \di g)^{\wedge 3}\, ,
\label{wzw}
\end{equation}
where $\mathcal{S} = \partial \mathcal{B}$. The inverse coupling constant $k$ sets the level of the 
$\widehat{\mathfrak{su}(2)}$ affine algebra. This action is conveniently written in terms of $SU(2)$ Euler angles, using 
\begin{equation}
g  = \mathrm{e}^{\frac{i}{2}\sigma_3 \psi_{\mathrm{L}}} \mathrm{e}^{\frac{i}{2}\sigma_1 \theta} \mathrm{e}^{\frac{i}{2}\sigma_3 \psi_{\mathrm{R}}}\, .
\label{euler}
\end{equation} 
Our strategy is now to supplement the \textsc{wzw} action and the linear dilaton yielding the background (\ref{chs-het}) with an 
Abelian right-moving current at level $k_g$  from the $SO(32)$ gauge sector of the heterotic string. The latter can 
be written in terms of a free (chiral) boson as $\bar{\jmath}_g = i \sqrt{k_g/\alpha'} \bar{\partial} X$.
The bosonic part of the worldsheet action reads:
\begin{multline}\label{action-het0}
S_{\textsc{het},0}=\frac{1}{2\pi\alpha'}\int\!\! \di^2 z\,  \Big[
\partial X^\mu \bar \partial X_\mu +\\
\frac{\alpha'k}{4}\big(\p \rho\,\pb \rho + \p\theta\,\pb\theta + \p\psi_{\mathrm{L}}\,\pb\psi_{\mathrm{L}} + 
\p\psi_{\mathrm{R}}\,\pb\psi_{\mathrm{R}} + 2\cos\theta \,\p \psi_{\mathrm{L}}\,\pb\psi_{\mathrm{R}} \big) 
+\p X \,\pb X \Big]\,,
\end{multline}
with the dilaton supported by the direction $\rho$.

A non-trivial gauge field in target space can be added as a current--current ``magnetic'' deformation:
\begin{equation}\label{def-action-het}
S_{\textsc{het}}=S_{\textsc{het},0} - \frac{1}{\sqrt{2k_g}\pi}\,\mu \int \di^2 z\, (j^3+:\psi^+\psi^-\!:)\bar\jmath^g\,,
\end{equation}
where $j^3(z)$ is a bosonic $SU(2)_{k-2}$ current\footnote{which is normalized as $j^3=i\sqrt{k}\big(\p\psi_{\mathrm{R}}+\cos\theta\,\p\psi_{\mathrm{L}}\big)$}, which is corrected by the fermion bilinear 
${:\psi^+\psi^-:}$ 
in the $\mathcal{N}=(1,0)$ \textsc{wzw} model. 
The perturbation (\ref{def-action-het}) has been studied extensively in~\cite{Israel:2004vv, Israel:2004cd}, following 
an earlier work~\cite{Kiritsis:1994ta}. As the marginal deformation is integrable, one gets a solvable theory 
for any finite value of $\mu$ in the range $[0,1/\sqrt{2})$. The background fields, that can be conveniently 
found by Kaluza--Klein reduction along $\partial_X$, read:
\begin{subequations}\label{dualTX}
\begin{align}
\di s^2 &= {\ds \frac{\alpha'k}{4}\Big(\di \rho^2+\mathrm{d}\theta^2+
\sin^2 \theta\, \mathrm{d}\psi_{\mathrm{L}}^2+(1-2\mu^2)(\mathrm{d}\psi_{\mathrm{R}}+\cos\theta\,\mathrm{d}\psi_{\mathrm{L}})^2\Big)\,,} \\
{\mathcal B}_{[2]} &= {\ds -\frac{\alpha'k}{4}\,\cos\theta\,\mathrm{d}\psi_{\mathrm{L}}\wedge \mathrm{d}\psi_{\mathrm{R}}\,,}\\
{\mathcal A}_{[1]}& = {\ds \sqrt{\frac{2k}{k_g}}\mu\,(\mathrm{d}\psi_{\mathrm{R}}+\cos\theta\,\di\psi_{\mathrm{L}})\,.}
\end{align}
\end{subequations}
The deformed geometry is thus a squashed $S^3$. Viewing the three-sphere as a Hopf fibration of an $S^1$ over an $S^2$ base, the 
backreaction of the magnetic field changes continuously the radius of the $S^1$ fibre along the line of deformation, up to  
$\mu=1/\sqrt{2}$ where the fiber degenerates. 

\subsection{Dynamical promotion of the current--current deformation}

The above supergravity solution~(\ref{dualTX}) is by construction an exact \textsc{cft}. Although it cannot be identified with the 
Abelian bundle over Eguchi--Hanson~(\ref{sugrasol}), it correctly reproduces some features of it. For a given value of the 
radial coordinate $\rho$ in (\ref{sugrasol}), one has a transverse three-sphere, squashed as 
in~(\ref{dualTX}), with an \textsc{ns-ns} two-form flux 
and a magnetic gauge field.  One can then attempt to get the string sigma-model for the 
background~(\ref{sugrasol}) as the dynamical promotion of the 
current--current marginal deformation~(\ref{def-action-het}) leading to (\ref{dualTX}). This  provides us with a 
first correspondence between a worldsheet conformal field theory and the resolution of orbifold singularities. 

The natural candidate coordinate for the promotion of the deformation in eq.(\ref{def-action-het}) is  $\rho$, i.e. 
the radial coordinate in the fivebranes geometry, originally supporting the dilaton. The constant deformation parameter of the background (\ref{dualTX}) is now promoted to a function of $\rho$ 
\begin{equation}
\mu=\mu(\rho)\, .
\end{equation} 
The formerly linear dilaton acquires a more complicated $\rho$-dependence, that we parametrize as:
\begin{equation}
\label{dilat-ansatz}
\Phi(\rho)
=-\tfrac{1}{2}\ln f(\rho)\,,
\end{equation}
the range of $\rho$ being tied to the boundary condition on $\mu$ as determined from the constant deformation. We can now determine the beta-function equations, to leading order in $\alpha'$.
\begin{itemize}
\item The Einstein equations (\ref{het1}) reduce to the two independent equations (where $'\equiv \p_{\rho}$):
\begin{subequations}
\label{eq-asym}
\begin{align}
{\ds \frac{f'}{f}+\frac{\mu''}{\mu'} + \frac{2\mu\mu'}{1-2\mu^2}}&= 0\,,\label{eq-asyma}\\[3pt] 
{\ds \left(\frac{f'}{f}\right)^2 - \frac{f''}{f}-\left(\frac{2\mu\mu'}{1-2\mu^2}\right)\frac{f'}{f}}&= 0\,,\label{eq-asymb}
\end{align}
\end{subequations}
\item The $B$-field equation~(\ref{het2}) is automatically satisfied by a general ansatz of the form $\mathcal{H}_{[3]}~=h(\rho)\,\Omega_3$, where $\Omega_3$ is the volume form on $S^3$, regardless of the value of $h(\rho)$.
\item The gauge field equation~(\ref{het3})  reproduces (\ref{eq-asym}a).
\item Finally, the vanishing of the  beta-function for the dilaton (\ref{het4}) yields, as usual, a constraint:
\begin{equation}
\label{constr-dilat}
\left(\frac{2\mu\mu'}{1-2\mu^2}\right)\frac{f'}{f} + \frac{2\mu'^2}{1-2\mu^2} - \frac{f''}{f}= -\frac{k}{6}\,\delta c
\end{equation}
relating $\mu$ and $f$ to the deficit of central charge $\delta c$, given in terms of $c_{SU(2)_k}=3\left(\frac{k-2}{k}\right)$ as (we focus on the critical case $c=3$):
\begin{equation}\label{eq:ccsol}
  \delta c ={ c - 1 - c_{SU(2)_k}}= {\frac{6}{k}}\, .
\end{equation}
\end{itemize}

\noindent
The system of equations (\ref{eq-asym}) can be recast in a more handy form by the following reparametrization:
\begin{equation}\label{change-var}
\nu(\rho)=\sqrt{1-2\mu(\rho)^2}\,.
\end{equation}
After integrating once, the system  becomes first-order. One can define a linear combination of eqs.~(\ref{eq-asyma},b) which decouples $\nu$ from $f$:
\begin{equation}
\label{eq-simple}
\frac{\partial}{\partial \rho}\ln\left(-\frac{\nu'}{\sqrt{1-\nu^2}}\right)-\kappa_1 \,\nu  = 0\,.
\end{equation}
This equation is exactly solved by the following (exhaustive) three-parameter family of functions:
\begin{equation}
\label{sol-y}
\nu(\rho) = \kappa_2 \sinh\big[\kappa_2(\rho+\delta)\big]\,\left(\frac{\sqrt{\kappa_2^2-\kappa_1^2}
-\kappa_1 \cosh\big[\kappa_2(\rho+\delta)\big]}{\kappa_2^2+\kappa_1^2\sinh^2\big[\kappa_2(\rho+\delta)\big]}\right)\,.
\end{equation}

The various constants can be determined by demanding that 
the function $\nu(\rho)$ must satisfy the same boundary conditions  as the constant deformation $\mu$ it derives from. These are  
\begin{equation}
\label{cond}
\lim_{\rho\rightarrow 0}\nu(\rho)=0\quad , \qquad 
\lim_{\rho\rightarrow \infty}\nu(\rho)=1\, .
\end{equation}
The former sets $\delta=0$, and the latter $\kappa_1=-\kappa_2=\kappa$. 
This selects the  solution $\nu(\rho)=\tanh(\kappa \rho)$, 
translating into
\begin{equation}\label{solution-y}
\mu(\rho)=\frac{1}{\sqrt{2}\cosh(\kappa \rho)}
\end{equation}
with $\rho\in [0,\infty[$. With this in hand, the second combination of eqs.~(\ref{eq-asyma},b) simplifies considerably:
\begin{equation}
\ds \frac{\p\ln f}{\p \rho}+\kappa \nu = 0\,,
\end{equation}
and gives the following non-trivial dilaton profile:
\begin{equation}\label{dilprof}
f(\rho)=\ds \mathrm{e}^{-2\Phi_0}\,\cosh(\kappa \rho)\,.
\end{equation}
Finally, eq.(\ref{constr-dilat}) can be rewritten in terms of $\nu$ only:
\begin{equation}\label{constr-y}
(\kappa \nu)^2 -2\kappa \nu' -\frac{\nu'^2}{1-\nu^2}=\frac{k}{6}\,\delta c\, ,
\end{equation}
which leaves $\Phi_0$ unconstrained as expected, while for the solution (\ref{solution-y}) it fixes $\kappa^2=1$, at the critical dimension.

The background generated by the dynamical asymmetric (magnetic) deformation of a configuration of 
heterotic fivebranes is obtained by replacing in~(\ref{dualTX}) the constant $\mu$ with 
the function $\mu (\rho)$ given in~(\ref{solution-y}), supplemented with the dilaton~(\ref{dilprof}). 
Setting $k_g=2$ and $k=2\ell^2$, we recognize the supergravity solution for the  $U(1)$ gauge bundle 
over Eguchi--Hanson space (\ref{sugrasol}). This assignment for the level of the $SU(2)$ will later acquire a precise significance in terms of anomaly cancellation for the corresponding gauged \textsc{wzw} model.  The 
$\mathbb{Z}_2$ orbifold of the three-sphere, which was not instrumental in the above analysis, will also have a precise \textsc{cft}  interpretation.

\section{A Coset CFT for the Heterotic Gauge Bundle}
\label{seccft}
In this section we consider the worldsheet conformal field theory description of the Abelian bundle over Eguchi--Hanson space, 
given by the heterotic supergravity solution~(\ref{sugrasol}). The sigma-model analysis done in 
sec.~\ref{secsigma} suggests that, from the worldsheet perspective,  
of the solution can be viewed as a {\it dynamical} current-current deformation of the $SU(2)$ 
\textsc{wzw} model. We find here the corresponding exact \textsc{cft} and analyze its spectrum.

\subsection{Heterotic fivebranes on an orbifold}
\label{blowdown}
As discussed in sec.~\ref{secsigma}, from the worldsheet \textsc{cft} point of view it is easier to take as a starting point the
 blow-down limit of  Eguchi--Hanson with the Abelian bundle turned on, i.e. the near-horizon geometry 
for  coincident heterotic fivebranes transverse to an A$_1$ singularity.  This is 
is given on the worldsheet by a $\mathbb{Z}_2$ orbifold of the  \textsc{chs} solution:
\begin{equation}
\mathbb{R}^{5,1} \times \mathbb{R}_\mathcal{Q} \times SU(2)_{k}/\mathbb{Z}_2\, ,
\label{chs}
\end{equation} 
For definiteness we choose the $Spin(32)/\mathbb{Z}_2$ heterotic string. 
This $\mathcal{N}=(1,0)$ worldsheet superconformal field theory is made of the following constituents:
\begin{itemize}
\item free $\mathcal{N}=(1,0)$ superfields $(X^\mu,\psi^\mu)$, $\mu=0,\ldots,5$ for the flat space-time part 
\item an $\mathcal{N}=(1,0)$ super-linear dilaton $(\varrho,\psi^\varrho)$ of background charge $\mathcal{Q}=\sqrt{\nicefrac{2}{\alpha' k}}$
\item  an $\mathcal{N}=(1,0)$ $SU(2)$ \textsc{wzw} model  at level $k$, corresponding  to a three-sphere of radius $\sqrt{\alpha 'k}$. 
The left-moving currents and 
their superpartners are denoted by $(J^\alpha,\psi^\alpha)$, $\alpha=1,\ldots,3$, forming an 
$\widehat{\mathfrak{su}(2)}$ super-affine algebra 
at level $k$. The right moving currents are denoted by $\bar{\jmath}^\alpha (\bar z)$, forming an
$\widehat{\mathfrak{su}(2)}$ affine algebra at level $k-2$ 
\item 32 right-moving Majorana-Weyl fermions from the gauge sector $\bar{\xi}^i$, with common spin structure
\item an $\mathcal{N}=(1,0)$ super-ghost system
\end{itemize}
The SU(2) \textsc{wzw} model is modded out by the $\mathbb{Z}_2$ action $\mathcal{I}~:g \mapsto -g$
that leaves invariant the current algebra. Consistency of the \textsc{cft} requires $k$ to be even. In order to keep $\mathcal{N=}(1,0)$ spacetime supersymmetry in six dimensions, one lets the orbifold act on the right-movers, i.e. on 
representations of the purely bosonic $\widehat{\mathfrak{su}(2)}$ affine algebra at level $k-2$.

One also considers an orbifold action in the $Spin(32)/\mathbb{Z}_2$ lattice, i.e. a non-trivial holonomy 
for the gauge field, that corresponds 
to the degeneration of the Abelian gauge bundle of shift vector $\vec{\ell}$ to a point-like Abelian instanton, see~\cite{Gremm:1999hm}. 
Using the fermionic representation of the $Spin(32)/\mathbb{Z}_2$ lattice, 
the orbifold sector $(\gamma,\delta)$, with $\gamma,\delta \in \mathbb{Z}_2$,  gives the affine character: 
\begin{equation}
\prod_{i=1}^{16} \vartheta \oao{u+2\ell_i \gamma}{v+2\ell_i \delta} (\bar \tau) \, ,
\end{equation}
where $(u,v)$ corresponds to the spin structure on the torus of modulus $\tau$, and 
$\vartheta \oao{u}{v} (\tau)$ to the usual free-fermion theta-function. In the following 
we choose to consider only bundles with vector structure.  Then the shifted lattice can be rewritten in a much simpler form using 
the periodicity of the theta-functions: 
\begin{equation}
\prod_{i=1}^{16} \vartheta \oao{u+2\ell_i \gamma}{v+2\ell_i \delta} (\bar \tau) = (-)^{\delta u \sum_i \ell_i}\,  \vartheta \oao{u}{v}^{16}(\bar \tau).
\label{ferbundle}
\end{equation}
Therefore, the only effect of the orbifold at the level of the partition function 
could be to flip the \textsc{gso} projection in the Ramond sector, but only in the case where $\sum_i \ell_i$ is odd. We will 
see below that $\sum_i \ell_i$ needs to be even for consistency of the theory. The twisted sector of the orbifold, 
given by $\gamma=1$, corresponds merely to a redefinition of the Neveu-Schwarz vacuum. This has no effect on the physical spectrum. 

Having understood its $Spin(32)/\mathbb{Z}_2$ part, one obtains the full partition function of the worldsheet \textsc{cft} as follows:
\begin{multline}
Z(\tau)= \frac{1}{(4\pi^2 \alpha' \tau_2)^3}\frac{1}{\eta^4 (\tau) \eta^4 (\bar \tau)} 
\int_0^\infty \!\! {\rm d}p \, \frac{(q\bar q)^{\frac{p^2}{k}}}{\eta (\tau )  \eta (\bar \tau)}\, \ \times \\ \times \ 
\frac{1}{2} \sum_{\gamma,\delta=0}^1 \sum_{2j=0}^{k-2} (-)^{\delta(2j+ (k/2-1)\gamma)} 
\chi^j ( \tau)\chi^{j+\gamma(k/2-2j-1)}  (\bar \tau)\, \ \times \\ \times \
\frac12 \sum_{a,b=0}^1 (-)^{a+b} \frac{\vartheta \oao{a}{b}^4 ( \tau)}{\eta^4 (\tau)}\,
 \frac12 \sum_{u,v=0}^1 (-)^{\delta u  \sum_i \ell_i}
\frac{\vartheta\oao{u}{v}^{16} (\bar \tau)}{\eta^{16} (\bar \tau)}\, ,
\label{partorb}
\end{multline} 
where $\chi^j (\tau)$ are affine $\widehat{\mathfrak{su}(2)}$ characters at level $k-2$ and $p$ denotes the momentum 
for the linear dilaton $\varrho$. 
Eight spacetime supercharges, built from the left-moving free fermions, are preserved by this background. 
This singular fivebrane solution actually preserves an $SO(32)$ gauge symmetry, 
as all the gauge currents $\bar{\xi}^i \bar {\xi}^j$ are invariant under the orbifold action.  In the relevant 
case $\sum \ell_i$ even, the full $Spin(32)/\mathbb{Z}_2$ is preserved.  The relation between the level $k$ of the supersymmetric $\widehat{\mathfrak{su}(2)}$ affine algebra and the fivebrane charge can be read from our supergravity 
analysis of sec.~\ref{secsugra}. Anticipating a little bit, we see that combining relation~(\ref{tad}) relevant to the near-horizon limit described here,
together with eq.~(\ref{quantn}), which will be discussed shortly, gives
\begin{equation}
Q_5|_{\text{n.h.}} = \frac{k}{2}-3 \, ,
\label{fivenum}
\end{equation}
assuming that~(\ref{tad}) is not modified by $\alpha'$ corrections.  

Non-singular points in the moduli space of the fivebranes are constructed by 
adding $\mathcal{N}=(4,4)$ marginal operators to the worldsheet action,, 
for instance the  $\mathcal{N}=2$ Liouville interaction, corresponding to evenly spaced fivebranes 
on a circle~\cite{Giveon:1999px}, modded out by a $\mathbb{Z}_2$ orbifold in the present case. 
The original $SO(4)_1$ affine symmetry of these fermions is broken to $SU(2)_1$ corresponding to the R-symmetry of the $\mathcal{N}=4$ 
algebra~\cite{Antoniadis:1994sr,Gava:2001gv,Murthy:2006eg}. 
The other fermions remain free, forming  an $SO(28)$ algebra. Overall one has an unbroken $SO(28)\times SU(2)$ gauge 
symmetry, enhanced to $SO(32)$ at the singularity.

\subsection{Eguchi--Hanson resolution as a gauged WZW model} 
We now investigate how to describe the resolved singularity~(\ref{sugrasol}) from the  conformal field theory (rather 
than sigma-model) perspective. Following the discussion of sec.~\ref{secsigma}, 
one would like to consider a normalizable marginal deformation of the 
heterotic fivebrane worldsheet \textsc{cft}~(\ref{chs}) of the form:
\begin{equation}
\delta S \sim \int {\rm d}^2 z\, \mu(\varrho) \, J^3 \bar{\jmath}^g \, ,
\label{dyndef}
\end{equation}
up to possible fermionic corrections, by taking a current
from the Cartan of the gauge group (e.g. $\bar{\jmath}^g = i :\!\bar{\xi}^3 \bar{\xi}^{4}\!:$ in the fermionic 
representation). The  radial function $\mu (\varrho)$ should decay fast enough for $\varrho \to \infty$, i.e. far from the fivebranes. 
In sec.~\ref{secsigma} we solved the sigma-model conformal equations and found a suitable explicit solution.  

Based on previous experience in a type \textsc{ii} context~\cite{Israel:2004ir}, a \textsc{cft} realization 
of this {\it dynamical deformation} can be found by replacing the linear dilaton $\varrho$ by  an auxiliary 
$SL(2,\mathbb{R})_{k+2}$ \textsc{wzw} model, where $\varrho$ is now the radial direction in AdS$_3$ global  
coordinates.\footnote{up to a normalization factor: $\varrho = \sqrt{\tfrac{\alpha' k}{4}}\, \rho$,
where the AdS$_3$ coordinate $\rho$ is defined below} The prescription is then to gauge  the currents 
$(J^3,\bar{\jmath}^g)$ used in the deformation~(\ref{dyndef}) together 
with the elliptic Cartan sub-algebra $(K^3,\bar{k}^3)$ of $\widehat{\mathfrak{sl}(2)}$. The $\varrho$ dependence of the 
AdS$_3$ metric gives in the gauged \textsc{wzw} model the dynamical factor needed in eq.~(\ref{dyndef}).

In heterotic coset constructions, while the holomorphic part of the theory needs to have (at least) 
$\mathcal{N}=1$ superconformal symmetry, freedom is left for the amount of supersymmetry on the 
anti-holomorphic bosonic side. 

\paragraph{(1,0) gauged WZW model} 
On can first  start with an 
$SU(2)_{k} \times \slr_{k'}$ $\mathcal{N}=(1,0)$ super-\textsc{wzw} model. Calling $(g,g')$ a corresponding 
group element, we gauge a $U(1)_\textsc{l}\times U(1)_\textsc{r}$ subgroup, parametrized by $\alpha$ and $\beta$:
\begin{equation}
(g,g')\to (\mathrm{e}^{i\sigma_3 \alpha}g\,,\, \mathrm{e}^{i\sigma_3 \alpha} g' \mathrm{e}^{i\sigma_3  \beta} \, 
) \, , 
\label{gauging}
\end{equation}
defining a $\mathcal{N}=(1,0)$ gauged \textsc{wzw} model. We observe that the two gauged $U(1)$ factors are {\it chiral}, 
i.e. one acts only on the left-movers and the second one only on the right-movers. 
The left-moving gauging  is anomaly-free provided that the levels of the $\widehat{\mathfrak{su}(2)}$ 
and $\widehat{\mathfrak{sl}(2,\mathbb{R})}$ super-affine algebras are the same, 
i.e. $k=k'$. The anomaly from the right gauging (parametrized by $\beta$) can by cancelled by a minimal coupling 
of the gauge field to (at least) one right-moving Weyl fermion, whose integral charge is labelled as $\ell$.  
The right-moving part of the  $\slr$ \textsc{wzw} model is purely bosonic. 
It defines an $\widehat{\mathfrak{sl}(2,\mathbb{R})}$ affine algebra at level $k+2$.
Therefore one gets an overall anomaly-free coset for 
\begin{equation}
k =  k'=2 (\ell^2-1)  \quad , \qquad \ell \in \mathbb{Z}\, .
\label{quantbis}
\end{equation}
The superconformal symmetry of this \textsc{cft} is actually enhanced to $\mathcal{N}=(4,0)$, see sec.~\ref{nonpert}. As we will see shortly, 
it corresponds to a single Abelian bundle of charge $\ell$ over \textsc{eh} space, of the kind 
discussed in sec.~\ref{secsugra}. In the blow-down limit, one finds the fivebrane solution whose partition function is given by~(\ref{partorb}), with a shift vector 
$\vec{\ell}=(\ell,0,\ldots,0)$.

\paragraph{(1,1) Gauged WZW model}
Another consistent model can be obtained by requiring also $\mathcal{N}=1$ superconformal symmetry for the right-movers. 
This can be achieved by pairing the $\mathcal{N}=(1,0)$ \textsc{wzw} models involved in the coset with the  
free right-moving fermions requested for $\mathcal{N}=(1,1)$ supersymmetry.  In the case discussed here one starts with an 
$SU(2)_{k} \times \slr_{k'}$ $\mathcal{N}=(1,1)$ super-\textsc{wzw} model in order to define a $\mathcal{N}=(1,1)$ 
coset \textsc{cft}. One also couples minimally an extra right-moving Weyl fermion with charge $\ell$.\footnote{
The whole coset construction can be recast as an $\mathcal{N}=(1,1)$ gauged \textsc{wzw} model as follows. The 
$\mathcal{N}=1$ super-\textsc{wzw} model  $SU(2)$ at level 2  consists only of three free fermions, 
as the purely bosonic affine algebra is trivial. Thus one can start with a right-moving 
super-$\widehat{\mathfrak{su}(2)}_2$  affine algebra, taking a Weyl fermion from the gauge sector together 
with a spectator Majorana-Weyl fermion. Gauging the Cartan subalgebra, one obtains a 
minimal coupling of the Weyl fermion to the gauge field.} In this second example the coset is anomaly free for
\begin{equation}
k = k'= 2 \ell^2  \quad , \qquad \ell \in \mathbb{Z}\, .
\label{quant}
\end{equation}
As we will see below, the superconformal symmetry of the coset is enhanced to $\mathcal{N}=(4,1)$. Adding 
three extra spectator right-moving Majorana-Weyl fermions, this symmetry is even enhanced to  $\mathcal{N}=(4,2)$.\footnote{Note  that for the smallest possible
charge $\ell=\pm1$, the coset is actually identical to the \textsc{cft} for a pair of separated heterotic fivebranes, discussed 
e.g. in~\cite{Murthy:2006eg}, putting aside the $\mathbb{Z}_2$ orbifold. This coset has 
$\mathcal{N}=(4,4)$ superconformal symmetry.} As the right-moving fermions of the 
$SU(2)_k$ super-\textsc{wzw} model remain free after this gauging, one has in total four right-moving 
interacting fermions $\bar{\xi}^{a}$, $a=1,\ldots 4$.

\subsubsection*{Lagrangian formulation}
Finding the background fields corresponding to a heterotic coset is more tricky than for the usual bosonic 
or type \textsc{ii} cosets, because of the worldsheet anomalies generated by the asymmetric gauging. We will follow closely the method used in~\cite{Johnson:1994jw,Johnson:2004zq}. 
A convenient way to find the metric, Kalb-Ramond  and gauge field backgrounds from the gauged \textsc{wzw} model is to bosonize the fermions 
before integrating out the gauge field. One needs eventually to refermionize in order to get a heterotic sigma-model in the standard 
form, i.e. (see~\cite{Sen:1985eb}):
\begin{equation}
S= \frac{1}{4\pi} \int\!\! \di^2 z\, \Big[ \tfrac{2}{\alpha'} (g_{ij} +\mathcal{B}_{ij}) \p X^i \pb X^j 
+ g_{ij} \psi^i \bar{\nabla} (\Omega_+) \psi^j +  \bar{\xi}^A \nabla(\mathcal{A})_{AB} \bar{\xi}^B
+\tfrac{1}{4}   \mathcal{F}^{AB}_{ij} \bar{\xi}_A \bar{\xi}_B \psi^i \psi^j \Big]
\label{sigmaaction}
\end{equation}
where the worldsheet derivative $\bar{\nabla}  (\Omega_+)$ is defined with respect to the spin connexion 
$\Omega_+$ on the conformal Eguchi--Hanson space with torsion (see app.~\ref{appsigma} for more details) and the derivative $\nabla (\mathcal{A})$ 
with respect to the gauge connexion $\mathcal{A}$.

For definiteness we choose to present the results for the the $(4,0)$ coset, with a single Abelian bundle of charge $\ell$. 
As for $SU(2)$, see eq.~(\ref{euler}), we parametrize 
the $\slr$ group elements in terms of Euler angles as 
\begin{equation}
g' =  \mathrm{e}^{\frac{i}{2}\sigma_3 \phi_{\mathrm{L}}} \mathrm{e}^{\frac{1}{2}\sigma_1 \rho} \mathrm{e}^{\frac{i}{2}\sigma_3 \phi_{\mathrm{R}}} \, .
\end{equation}
The gauged \textsc{wzw} model resulting from the asymmetric gauging~(\ref{gauging}) reads
\begin{multline}\label{sigm-gaug}
S(A)= S_{SU(2)}\,+\,S_{\slr} +S_\text{Fer} (A) \\
+\,\frac{1}{8\pi}\int \di^2 z\, \Big[ 2 i A_1\sqrt{k+2}\,\bar k_3
+2i \bar A_2 \big(\sqrt{k-2}\,j_3+\sqrt{k+2}\,k_3\big) \\
- (k+2) \big(A_1 \bar A_1+A_2 \bar A_2+2\cosh \rho\, A_1\bar A_2\big) +(k-2) A_2 \bar A_2\Big]
\, ,
\end{multline}
where the first two terms are standard bosonic \textsc{wzw} actions for $SU(2)$ at level $k-2$ and
$\slr$ at level $k+2$ respectively, see eq.~(\ref{wzw}). We also introduced the left- and right-moving bosonic $\slr$ currents normalized as $k_3=i\sqrt{k\text{+}2}\,(\partial \phi_{\mathrm{R}}+\cosh \rho\,\partial \phi_{\mathrm{L}})$ and $\bar k_3=i\sqrt{k\text{+}2}\,(\bar\partial \phi_{\mathrm{L}}+\cosh \rho\,\bar\partial \phi_{\mathrm{R}})$. Using the on-shell gauge
invariance one can enforce the gauge-fixing condition $\phi_{\mathrm{L}}=\phi_{\mathrm{R}}=0$. 
The third term $S_\text{Fer} (A)$ in eq.~(\ref{sigm-gaug}) is the action for the worldsheet fermions, 
that uses covariant derivatives w.r.t. the worldsheet gauge 
fields. It includes terms for the left-moving Majorana-Weyl fermions $\psi^{1,2}$ and $\lambda^{1,2}$ from the $SU(2)$ 
and $\slr$ super-\textsc{wzw} models respectively, together with four right-moving Majorana-Weyl fermions\footnote{Only two 
of these four right-moving fermions will be interacting at the end of the calculation. However four of them are needed in order to bosonize 
fully the left- and right-moving worldsheet fermions.} $\bar{\xi}^a$, $a=1,\ldots,4$. 
One gets:  
\begin{multline}
S_\text{Fer} (A)= \frac{1}{4\pi} \int \di^2 z \Big[
\psi^1 \db \psi^1 + \psi^2 \db \psi^2 
+\lambda^1 \db \lambda^1 + \lambda^2 \db \lambda^2 -2 \bar{A}_2 (\psi^1  \psi^2  +\lambda^1  \lambda^2)\\
+\sum_{a=1}^4 \bar{\xi}^a \d \bar{\xi}^a -2 \ell A_1  \bar{\xi}^3  \bar{\xi}^4
\Big]\, .
\end{multline}

In this coset the classical anomalies from the bosonic gauged \textsc{wzw} model are compensated by the
quantum anomalies from the chiral fermions, provided that $k=2\ell^2-2$.\footnote{Mind that before integrating out the gauge fields, we are working at finite level $k$, so that the above relation is dictated by anomaly cancellation for the {\it bosonic} level. After solving for the gauge fields, we can take the infinite $k$ limit, which gives rise to the relation (\ref{quant}).}
The next step is to bosonize the fermions involved in the coset in terms of a pair of 
canonically normalized free bosons $\Phi_{1,2}$ compactified at the fermionic radius, i.e.
\begin{subequations}
\begin{align}
\partial \Phi_1 = \, :\! \psi^1 \psi^2\!:\,, & \qquad
\bar \partial \Phi_1 = \, :\! \bar \xi^1  \bar \xi^2\!:\,, \\
\partial \Phi_2 = \, :\! \lambda^1 \lambda^2\!:\,, & \qquad
\bar \partial \Phi_2 = \, :\! \bar{\xi}^3  \bar{\xi}^4\!:\,.
\end{align}
\end{subequations} 
Properly taking into account the anomalies (see~\cite{Johnson:1994jw}) we arrive at the action
\begin{multline}
S(A)= \frac{k+2}{8\pi} \int {\rm d}^2 z \, \d \rho \db \rho  +\frac{k-2}{8\pi} \int {\rm d}^2 z \Big[
\d \theta \db \theta + \d \psi_{\mathrm{L}} \db \psi_{\mathrm{L}} + \d \psi_{\mathrm{R}} \db \psi_{\mathrm{R}} + 2 \cos \theta \, \d \psi_{\mathrm{L}} 
\db \psi_{\mathrm{R}}\Big]\\
 +\frac{1}{4\pi} \int \!\! d^2 z\, (\p \Phi_1 \pb \Phi_1+ \p \Phi_2 \pb \Phi_2)\\
-\,\frac{1}{2\pi}\int \!\! d^2 z\, \Big[ \bar A_2 
\left( \tfrac{k-2}{2}( \d \psi_{\mathrm{R}} + \cos \theta \, \d \psi_{\mathrm{L}})+\d \Phi_1+\d \Phi_2 \right)
\\+  A_1\, \ell \db \Phi_2
+ \tfrac{k+2}{2} \cosh \rho A_1\bar A_2 - \ell A_2 \bar A_1 \Big]\,.
\end{multline}
Considering  the large $k$ (i.e. large charges) limit , upon integrating out classically the gauge fields
the action reduces to:
\begin{multline}
\label{intsigma}
S= \frac{k}{8\pi} \int \!\!{\rm d}^2 z \,\Big[ \d \rho \db \rho  +
\d \theta \db \theta + \d \psi_{\mathrm{L}} \db \psi_{\mathrm{L}} + \d \psi_{\mathrm{R}} \db \psi_{\mathrm{R}} + 2 \cos \theta \, \d \psi_{\mathrm{L}} 
\db \psi_{\mathrm{R}}+
 \tfrac{2}{k} (\p \Phi_1 \pb \Phi_1+ \p \Phi_2 \pb \Phi_2)\Big]\\
+\,\frac{1}{2\pi}\int \!\! d^2 z\, \frac{1}{\cosh \rho}\big[
\d \psi_{\mathrm{R}} + \cos \theta \, \d \psi_{\mathrm{L}}+\tfrac{2}{k}(\d \Phi_1+\d \Phi_2 )\big] \ell\, \db \Phi_2\,.
\end{multline}
In order to fermionize back, one has to rewrite the terms involving the fields $\Phi_{1,2}$ 
in the action~(\ref{intsigma}) using a Kaluza-Klein-like form (see~\cite{Johnson:1994jw}):
\begin{align}
\p \Phi_1 \pb \Phi_1&+ \p \Phi_2 \pb \Phi_2+\frac{2\ell}{\cosh \rho}\big[
\d \psi_{\mathrm{R}} + \cos \theta \, \d \psi_{\mathrm{L}}+\tfrac{2}{k}(\d \Phi_1+\d \Phi_2 )\big]\db \Phi_2
\nonumber\\
&=\ \p \Phi_1 \pb \Phi_1+
\left|\partial \Phi_2 +\ell \frac{\d \psi_{\mathrm{R}} + \cos \theta \, \d \psi_{\mathrm{L}}}{\cosh \rho}\right|^2
\nonumber\\
&\quad+\frac{\ell}{\cosh \rho} \left[ (\d \psi_{\mathrm{R}} + \cos \theta \, \d \psi_{\mathrm{L}}) \db \Phi_2
-(\db \psi_{\mathrm{R}} + \cos \theta \, \db \psi_{\mathrm{L}}) \d \Phi_2\right]\nonumber\\
&\quad-\frac{\ell^2}{\cosh^2 \rho} |\d \psi_{\mathrm{R}} + \cos \theta \, \d \psi_{\mathrm{L}}|^2
+\frac{4\ell}{k\cosh \rho} (\d \Phi_1+\d \Phi_2 ) \db \Phi_2
\, ,
\end{align}
the last two terms giving a correction to the sigma-model metric and the four-fermion interaction respectively. 
Then after re-fermionization, one obtains a heterotic sigma-model of the form~(\ref{sigmaaction}), whose bosonic part reads:
\begin{multline}
S_\textsc{b}= \frac{k}{8\pi} \int \!\!{\rm d}^2 z \Big[ \d \rho \db \rho  +
\d \theta \db \theta + \sin^2 \theta \, \d \psi_{\mathrm{L}} \db \psi_{\mathrm{L}} + \tanh^2 \rho \, (\d \psi_{\mathrm{R}} +\cos \theta \, \d \psi_{\mathrm{L}} )(\db \psi_{\mathrm{R}} +\cos \theta \, \db \psi_{\mathrm{L}} )\\
+\cos \theta (\d \psi_{\mathrm{L}} \db \psi_{\mathrm{R}}-\d \psi_{\mathrm{R}} \db \psi_{\mathrm{L}})\Big] \,.
\label{bossigma}
\end{multline}
The fermionic part of the sigma-model action is (renaming the left-moving fermions $\lambda^{1,2}$ as $\psi^{3,4}$)
\begin{multline}
S_\textsc{f} = \frac{1}{4\pi} \int \!{\rm d}^2 z \,\Big\{\sum_{a=1}^4  \psi^a \db \psi^a +\bar \xi^1 \d \bar \xi^1 +
\bar \xi^2 \d \bar \xi^2 +\\  (\bar\xi^3\ ,\bar\xi^4) \left[ \mathbb{I}_2\,\d +\frac{\ell}{\cosh \rho}(\d \psi_{\mathrm{R}} +\cos \theta \, \d \psi_{\mathrm{L}} ) \,i\sigma^2
\right]\oaop{\bar \xi^3}{\bar \xi^4} +\frac{2}{\ell \cosh \rho} (\psi^1 \psi^2 + \psi^3 \psi^4) \bar \xi^3 \bar \xi^4
\Big\} .
\label{fersigma}
\end{multline}
In addition there is a dilaton coming from the integration of the worldsheet gauge fields. It reads:
\begin{equation}
\Phi = \Phi_0 - \tfrac{1}{2} \ln \cosh \rho\, .
\end{equation}
Adding the six-dimensional flat space part we recognize  the supergravity background~(\ref{sugrasol}) with a single $U(1)$ bundle of charge $\ell$.  
The dilaton zero-mode $\Phi_0$  is given in terms of the double scaling parameter by eq.~(\ref{dilateq}). As the 
Bianchi identity~(\ref{bianchi}) is not satisfied locally for this solution at any point, these background fields are expected 
to receive  perturbative $\alpha'$ corrections, despite the $(4,0)$ superconformal symmetry of the coset. 
Any potential corrections should  preserve the $SU(2)$ isometry corresponding to the right-moving $\widehat{\mathfrak{su}(2)}_{k-2}$ 
algebra. We did not attempt to compute explicitly these corrections (this can be done by using the methods of~\cite{Tseytlin:1992ri,Bars:1993zf},
since we are dealing with a coset \textsc{cft}).\footnote{In~\cite{Johnson:2004zq}
a Lorentzian analogue of this $\mathcal{N}=(1,0)$ heterotic coset was considered (case $\delta=0$ there) and the $1/k$ corrections 
to the background fields were computed. One may wonder whether the Wick rotation 
of these results is relevant for the background discussed here. It turns out not to be likely the case, 
because none of the two gauged $U(1)$ subgroups in the model of~\cite{Johnson:2004zq} acts chirally, while in our 
case the left-moving gauging is identical to the holomorphic side of the type II $(4,4)$ \textsc{cft} for 
fivebranes on a circle defined in~\cite{Israel:2004ir}. One can check that asymptotically (for large $\rho$) 
the full $SO(4)$ isometry of the solution is not restored there, whereas it has to be so in our model.}

\subsubsection*{More general bundles}
It is easy to generalize  the coset construction to the generic $U(1)^{16}$ bundle of the solution~(\ref{sugrasol}).
From the worldsheet point of view, one has to couple minimally 
the coset~(\ref{gauging}) to $16$ right-moving Weyl fermions instead of one.  
We recall that the embedding of the right gauging into the gauge group is then specified by a shift vector $\vec{\ell}$: 
\begin{equation}
\vec{\ell}=(\ell_1,\ell_2,\ldots,\ell_{16})  \ , \qquad  \ell_\alpha  \in \mathbb{Z} \, , \ \alpha = 1,\ldots,16  \qquad\text{or} \qquad \ell_\alpha \in \mathbb{Z} + \tfrac{1}{2} \, , \  \alpha = 1,\ldots,16  \, ,
\label{vecshift}
\end{equation}
for vectorial and spinorial shifts respectively. The quantization condition~(\ref{quant}) becomes:
\begin{equation}
k = 2 (\vec{\ell}^2-1) \,.
\label{quantn}
\end{equation}
This coset has $\mathcal{N}=(4,0)$ superconformal symmetry generically. 
The background fields can be found along the same lines as in the previous example. The only difference, 
appart from the relation~(\ref{quantn}) specifying $k$ as a function of the charges, lies in the presence of 
components of the Abelian gauge background along the whole Cartan subalgebra, instead of in only one generator (see sec.~\ref{secsugra}). 
The Abelian gauge field, embedded in  $SO(32)$, reads:
\begin{equation}
\mathcal{A}_{[1]} = \frac{1}{2\cosh \rho}(\di \psi_{\mathrm{R}} + \cos \theta \,\di \psi_{\mathrm{L}})
\, \vec{\ell} \cdot \vec{H}\, ,
\end{equation}
where $H^i$ are the generators of the Cartan. It allows to break a larger part of the gauge group at the $A_1$ singularity.

In a similar fashion one can generalize the $\mathcal{N}=(4,1)$ coset coming from the 
$\mathcal{N}=(1,1)$ gauged \textsc{wzw} model. It is nothing but a specific case of~(\ref{vecshift}), 
setting one of the charges to one, say $\ell_{1}=1$. It allows 
the $\slc$ part of the coset to have $\mathcal{N}=2$ right-moving superconformal symmetry.\footnote{To supersymmetrize the 
$SU(2)$ factor, one requires  three extra right-moving Majorana-Weyl fermions, but as they are free it will 
essentially have no effect.} 

Interestingly, this coset \textsc{cft}, setting $\ell_1=1$ and $\ell_i=0$ for $i>3$,  can be used in a type \textsc{ii} superstrings construction (as the  $\mathcal{N}=(1,1)$ 
superconformal symmetry is mandatory in that case). Turning to  type \textsc{ii}A or type \textsc{ii}B compactified 
on $T^2 \times K3$, we can consider a $U(1)^2$ Abelian bundle of the sort discussed here, where the gauge 
background of charges $(\ell_{2},\ell_{3})$ uses Kaluza-Klein gauge fields from the two-torus, i.e. it corresponds 
to a $T^2$ fibration over Eguchi--Hanson space. We plan to come back to these models in a companion paper.\footnote{Using 
simultaneously these Kaluza--Klein gauge fields and 'internal' gauge fields one can also construct similar fibered 
solutions in heterotic strings, similar to the models of~~\cite{Fu:2008ga}.}

\subsection{Algebraic construction of the heterotic spectrum}
We managed to identify the supergravity solution~(\ref{sugrasol}) for a an Abelian  bundle background over Eguchi--Hanson space
with a worldsheet \textsc{cft} consisting of a flat six-dimensional space--time tensored with a gauged \textsc{wzw} model. This will 
allow us to obtain the full string spectrum, using the standard coset construction. Because of its relevance 
in type \textsc{ii} and its simplicity, we will mainly discuss the $\mathcal{N}=(4,2)$ coset 
with the simple Abelian bundle of shift vector $\vec{\ell}=(1,\ell,0,\ldots,0)$. Generalization to the more generic case is rather 
straightforward. One considers then the $\mathcal{N}=(1,1)$ gauged \textsc{wzw} model 
\begin{equation}
\label{cftback}
\frac{SU(2)_k \times SL(2,\mathbb{R})_k \times SU(2)_{2,\, \textsc{r}}}{U(1)_\textsc{l} \times U(1)_\textsc{r}} \, ,
\end{equation}
the last factor corresponding to the superconformal affine algebra realized by the right-moving fermions. 
The asymmetric gauging of the product \textsc{wzw} model imposes the following zero-mode constraints on  physical states:
\begin{equation}
(J^3_0 -K^3_0)\, |\text{Phys}\rangle  =0 \quad , \qquad  
(\bar K^3_0 -\ell {\bar \jmath}^g_0)\, |\text{Phys}\rangle=0 \, .
\end{equation}
These constraints are  resolved by splitting the $SU(2)_k$ \textsc{wzw} model 
in terms of its super-coset and the $U(1)$ factor which is gauged, according to $SU(2)_k \sim SU(2)_k/U(1)\times U(1)_k$. 
Some details  about  $SU(2)/U(1)$ super-coset characters $C^j_m \oao{a}{b}$  are given in app.~\ref{appchar} (one does not need to decompose the right-moving characters, as the coset preserves 
the full $\widehat{\mathfrak{su}(2)}_{k-2}$ affine symmetry on the bosonic right-moving side). 
We introduce in addition $\slc$ (extended) characters ${\rm Ch} (J,M) \oao{a}{b}$ where $J$ is the $\slr$ spin, 
discussed also in appendix. In this 
$\mathcal{N}=(4,2)$ example, superconformal $\slc$ characters appears both for the left- and right-movers.

Starting from the partition function in the blow-down limit (\ref{partorb}) and plugging in 
the coset~(\ref{cftback}) instead of the $\mathbb{R}_\mathcal{Q} \times SU(2)$ \textsc{cft}, one obtains the partition function
for heterotic strings on Eguchi--Hanson with an Abelian bundle.  
It first consists in continuous representations of the $\slc$ super-coset (i.e. with $J = \tfrac{1}{2}+ip$):
\begin{multline} \label{partfunc}
Z_c(\tau)= \frac{1}{(4\pi^2 \alpha'\tau_2)^3}\frac{1}{\eta^4 \bar{\eta}^4} 
\, \frac12 \sum_{a,b=0}^1 (-)^{a+b}
\frac{\vartheta \oao{a}{b}^2}{\eta^2}\ \times
\\ \times \ 
\frac12 \sum_{\gamma,\delta=0}^1 \sum_{2j=0}^{k-2}(-)^{\delta(2j+(k/2-1)\gamma)} 
\sum_{m \in \mathbb{Z}_{2k}} 
C^j_m \oao{a}{b} \bar{\chi}^{j+\gamma(k/2-2j-1)} \frac12 \sum_{u,v=0}^1 
 (-)^{(\ell+1) \delta u}
\frac{\bar{\vartheta}\oao{u}{v}^{14}}{\bar{\eta}^{14}} \ \times \\ \times \ 
\int_0^\infty \!\! \text{d} p\,\, 
{\rm Ch}_c (\tfrac{1}{2}+ip,\tfrac{m}{2}) \oao{a}{b} 
\sum_{n \in \mathbb{Z}_{2\ell}} \mathrm{e}^{i\pi v (n+ \frac{u}{2}) }
\overline{\rm Ch}_c \big( \tfrac{1}{2}+ip,\ell(n+\tfrac{u}{2}) \big) \oao{u}{v} \, . 
\end{multline}
Actually, using the definition of continuous $\slc$ characters (see eq.~(\ref{extcontchar}) in app.~\ref{appchar}), this partition 
function is identical to the partition function in the blow-down limit~(\ref{partorb}). 
This is not surprising as the continuous spectrum corresponds to asymptotic 
states in the weakly coupled region $\varrho \to \infty$, i.e. far from the bolt of Eguchi--Hanson.

However, a set of new discrete states localized near the bolt appears after 
the resolution of the singularity. Built on discrete representations of $\slc$ (with spin 
$\tfrac{1}{2}<J<\tfrac{k+1}{2}$), they give a spectrum of localized states:
\begin{multline} 
Z_d(\tau)= \frac{1}{(4\pi^2 \alpha' \tau_2)^3}\frac{1}{\eta^4 \bar{\eta}^4} 
\, \frac12 \sum_{a,b=0}^1 (-)^{a+b}
\frac{\vartheta \oao{a}{b}^2}{\eta^2}\ \times
\\ \times \ 
\frac12\sum_{\gamma,\delta=0}^1 \sum_{2j=0}^{k-2}(-)^{\delta(2j+(k/2-1)\gamma)} 
\sum_{m \in \mathbb{Z}_{2k}} 
C^j_m \oao{a}{b} \bar{\chi}^{j+\gamma(k/2-2j-1)} \frac12 \sum_{u,v=0}^1
(-)^{(\ell+1) \delta u}
 \frac{\bar{\vartheta}\oao{u}{v}^{14}}{\bar{\eta}^{14}} \ \times \\ 
\sum_{2J=2}^{k}\!\!
{\rm Ch}_d (J,\tfrac{m}{2}-J-\tfrac{a}{2}) \oao{a}{b} 
\sum_{n \in \mathbb{Z}_{2\ell}}\!\! \mathrm{e}^{i\pi v (n+ \frac{u}{2}) }
\overline{\rm Ch}_d \big( J,\ell(n+\tfrac{u}{2})-J -\tfrac{u}{2}\big) \oao{u}{v} 
\delta_{2J-m+a,0}^{[2]}\, \delta_{2J-(\ell-1) u,0}^{[2]}
\label{discspec}
\end{multline}
with $\delta^{[2]}$ the mod-two Kronecker symbol. These discrete states break part of the gauge symmetry, as explained below.

One can read from this spectrum that the theory preserves $\mathcal{N}=(1,0)$ space-time supersymmetry 
in six dimensions, as the original singular solution~(\ref{chs}). We observe that, 
while the 'static' marginal deformation $\int \di^2 z \, J^3 \bar{\jmath}^g$ actually breaks 
all space-time supersymmetries, the 'dynamical' marginal deformation~(\ref{dyndef}) preserves 
all of them. 

\subsubsection*{Massless localized states}
\label{massless}
It is interesting to find the spectrum of massless string states whose wave-function is localized near the resolved singularity, 
in order to compare this \textsc{cft} description with the supergravity regime. These localized states are 
not expected on general grounds to comprise necessarily the full massless spectrum of the \textsc{ale} background, but only  those 
degrees of freedom that remain interacting in the double scaling limit of the two-cycle. 

Physics of point-like 
heterotic instantons sitting on K3 singularities is discussed in particular in~\cite{Aspinwall:1996vc}. 
In the blow-down limit, these degrees of freedom are
expected to be part of some  'Little String Theory'~\cite{Seiberg:1997zk} in six dimensions, 
of the sort discussed in~\cite{Blum:1997mm}. Appearance of massless tensor multiplets in the blow-down limit is responsible 
for the existence of non-trivial infrared fixed points~\cite{Seiberg:1996qx}. As for \textsc{ns5} branes in 
type \textsc{ii} theories~\cite{Aharony:1998ub}, the string background we consider here is expected to give a
holographic description of the latter.

These localized states belong to the 
discrete representations of $\slc$, whose contribution to the partition function is given by eq.~(\ref{discspec}). 
Massless bosonic states are obtained as \textsc{ns-ns} states of dimension $(\Delta=1/2,\bar\Delta= 1)$ in the 
heterotic worldsheet \textsc{cft}. Our strategy will be to first look at states of dimension $(1/2,1/2)$ 
as we would do in type \textsc{ii} superstrings. The left-moving part of the \textsc{cft} has 
$\mathcal{N}=4$ superconformal symmetry, while the right-moving part is made 
of an $\mathcal{N}=2$ $\slc$ part together with a bosonic  $SU(2)_{k-2}$ factor. Therefore one can study  
left and right chiral states of the $\mathcal{N}=4$ and $\mathcal{N}=2$ algebras respectively; the relevant chiral primaries are 
discussed in app.~\ref{appchar}. A physical state of dimension $(1/2,1)$ 
can be found from these 'chiral' states either by:
\begin{itemize}
\item adding a fermionic oscillator $\bar{\xi}^a_{-1/2}$  from the free gauge sector G (which is 
$SO(28)_1$ in this particular example). This gives a state in the fundamental  representation 
of G.
\item taking the right superconformal descendant of the $(1/2,1/2)$ state using the global 
right-moving superconformal algebra of the $\slc$ coset (i.e. acting with $\bar{G}_{-1/2}$). This
gives a singlet of G.
\end{itemize}
Note that the left-moving part of the theory is similar to the case of type \textsc{ii} fivebranes on a circle, 
discussed in detail in e.g.~\cite{Israel:2004ir,Aharony:2004xn}. It has extended $\mathcal{N}=4$ superconformal symmetry. 
We assume below that $\ell \geqslant 2$. In the $\ell\leqslant -2$ case the analysis is similar.\\ 

\noindent $\diamond$ {\it Left-moving part of massless operators}\\
On the supersymmetric left-moving side, an \textsc{ns}  primary operator of dimension $\Delta=1/2$ can be obtained 
first from a chiral operator of  $SU(2)_k/U(1)$ of conformal dimension $\Delta=\tfrac{1}{2}-\tfrac{j+1}{k}$, where $j$ is 
integer due to the orbifold constraint. It  can be tensored with a chiral primary of the super-coset $\slc$ of dimension 
$\Delta=\tfrac{J}{k}$, with spin $J=j+1$. One gets a \textsc{gso}-even state of dimension $\Delta=1/2$, that we denote $(c,c)$. 
A second possibility is to consider an anti-chiral operator of $SU(2)_k/U(1)$ of 
dimension $\Delta=\tfrac{j}{k}$  tensored  with an $\slc$ anti-chiral operator with 
$\Delta=\tfrac{1}{2}-\tfrac{J-1}{k}$, and again $J=j+1$. It gives also a physical state with $\Delta=1/2$, that we call $(a,a)$.\\

\noindent $\diamond$  {\it Right-moving part: untwisted sector}\\
On the bosonic side, one can first consider states in the untwisted sector of the $\mathbb{Z}_2$ orbifold, i.e. 
with $\gamma=0$. The $SU(2)$ and $\slr$ spins have to be the same as for the 
left-movers, and are related through $J=j+1$. Conformal primary states of this  \textsc{cft} 
are made of a  primary of bosonic $SU(2)_{k-2}$, of dimension $\bar \Delta = j(j+1)/k$, tensored with a 
primary of the $\slc$ super-coset. One can first choose a chiral primary with $\bar \Delta=\tfrac{j+1}{k}$. 
It gives a physical state of dimension $\bar \Delta =\tfrac{(j+1)^2}{2\ell^2}$, hence only the case 
 $j=\ell-1$ can give a  $(1/2,1/2)$ state, that we call $\bar{c}_u$.  A second possibility is 
to use an anti-chiral $\slc$ primary of dimension  $\bar \Delta =\tfrac{1}{2}-\tfrac{j}{k}$. The only 
state with $\bar \Delta=1/2$ is now obtained for $j=0$, and denoted $\bar{a}_u$. \\

\noindent $\diamond$ {\it Right-moving part: twisted sector}\\
The twisted sector of the $\mathbb{Z}_2$ orbifold, i.e. the states with $\gamma=1$ 
in the partition function~(\ref{discspec}), also gives discrete physical states. 
We make the assumption, which will be motivated later on, that $k/2$ is odd.  
In this case the right-moving $SU(2)$ spin reads $\tfrac{k}{2}-j-1$ in the twisted sector, 
while the $\slr$ spin is still $J=j+1$. One can first consider 
a chiral primary of $\slc$. A state with $\bar \Delta =1/2$ can be found only 
for $j=\ell^2-1$. We denote this state $\bar{c}_t$. It has \textsc{gso}-parity 
$(-)^\ell$. Using instead an anti-chiral primary of $\slc$, one finds a $(1/2,1/2)$ state only for 
$j=\ell (\ell-1)$. This state, denoted $\bar{a}_t$, has also \textsc{gso}-parity 
$(-)^\ell$. 
\vspace{0.5cm}

\begin{figure}[!!ht]
$$
\begin{array}{|l||c||l||c|c||c|}
\hline
\;SU(2)_{k-2}\, \text{spin} & \text{Left-moving} & \;\text{Right-moving}  &  \multicolumn{2}{|c||}{\;U(1)_{\bar{\textsc{r}}\;} \ \text{charge}}
& \;(-)^{\tilde{F}} \ \textsc{gso}\ \text{parity}\; \\
&& & \mathbf{1} & \Box &\\
\hline
\;j=\ell-1 & \;(a,a) + (c,c) \;& \;\bar{c}_u \ \text{(untwisted)} &\;\tfrac{1}{\ell}-1\;&\tfrac{1}{\ell} & 1\\
\;j=0 &\;(a,a) + (c,c) \;& \;\bar{a}_u \ \text{(untwisted)}&0&-1 & 1\\
\;j=\ell^2-1 &  \;(a,a) + (c,c) \;& \;\bar{c}_t\ \text{(twisted)} &0&1& \;(-)^{\ell+1}\;\\
\;j=\ell(\ell-1) &  \;(a,a) + (c,c) \;& \;\bar{a}_t\ \text{(twisted)} &\;1-\tfrac{1}{\ell}\;&-\tfrac{1}{\ell} &(-)^{\ell+1}\\[2pt]
\hline
\end{array}
$$
\caption{\it Spectrum of massless discrete states, built from left and right chiral/antichiral states ($\ell \geqslant 2$).
For each line one has one singlet ($\mathbf{1}$) and one fundamental ($\Box)$ hypermultiplets
w.r.t. the unbroken gauge group $\mathrm{G}$.}
\label{tabops}
\end{figure}

Using the method outlined in the beginning, one obtains for each of the states constructed above two $(1/2,1)$ states, one in the fundamental  and the other in the trivial representations of the gauge group. By worldsheet spectral flow one obtains in each case a half hypermultiplet of $\mathcal{N}=1$ supersymmetry in six dimensions. Putting
together the $(a,a)$ and $(c,c)$ states for given $SU(2)$ spin and G representation one obtains full hypermultiplets.

The massless spectrum that we obtain is summarized in tab.~\ref{tabops}. We also give the charge of the states under the $U(1)_{\bar{\textsc{r}}}$
right-moving worldsheet R-symmetry of the $\slc$ coset, that becomes an extra Abelian gauge symmetry. We see that, depending on the parity of $\ell$,
the twisted sector massless states can be \textsc{gso}-odd or \textsc{gso}-even. In the latter case,
which is the relevant one as we shall see, one has overall 4 singlet and 4 fundamental hypermultiplets, regardless of the value of $\ell$.

\subsubsection*{Gauge bosons}
Six-dimensional heterotic compactifications with gauge bundles have generically multiple massive
$U(1)$ factors, whose mass terms come from the generalized Green-Schwarz mechanism. Note that, unlike a 
four-dimensional compactification, there is no one-to-one correspondence between massless and anomalous $U(1)$ factors.\footnote{
An anomalous $U(1)$ stays massless if only  two-forms are involved in the \textsc{gs} mechanism, while a massive $U(1)$
can still be anomaly free if the sum of all \textsc{gs} diagrams vanishes, see e.g.~\cite{Honecker:2006dt}.} 

In the worldsheet \textsc{cft} description, a vector boson in spacetime can be obtained on the left-moving
side by tensoring  $\psi^\mu_{-1/2} |0\rangle_{\textsc{ns}}$ for the $\mathbb{R}^{5,1}$ part of the \textsc{cft} with an
$(a,c)$ state, made of an antichiral $SU(2)/U(1)$ and a chiral $\slc$ primaries, at equal spins $j=J$. The conformal
dimension of these \textsc{gso}-even states is generically $\Delta=1/2+2j/k$. The condition $j\geqslant 1$ guarantees that the wave-function of the 
operator is normalizable, for the $\slc$ part.  On the right-moving side, the vertex operator for the gauge boson used in the fibration
is obtained starting with an anti-chiral primary of $\slc$.  
Adding the $SU(2)_{k-2}$ contribution, one obtains also a conformal dimension $\bar \Delta=1/2+(j^2+1)/k$. Level-matching selects the spin
$j=1$. The superconformal  descendant of this state gives then the physical vertex operator for the massive Abelian gauge boson corresponding 
to the particular direction along which we turn on the  bundle in the Cartan of $SO(32)$.

One can straightforwardly generalize this analysis to a $U(1)^{16}$ bundle with generic charges. The mass of the Abelian 
gauge boson, whose direction in the Cartan is specified by the shift vector $\vec{\ell}$, is given as:
\begin{equation}
\mathfrak{m} = \frac{2}{\sqrt{\alpha'}} \, \frac{1}{\sqrt{\vec{\ell}^2-1}}.
\label{multimass}
\end{equation}

The affine currents constructed with the 32 fermions of the gauge sector, that commute with the Abelian affine current used for the gauge bundle, realize 
the affine algebra corresponding to the unbroken gauge group. The associated string states are obtained by taking on the right-moving side the identity of $\slc$ tensored with  states of the form $\bar \xi^i_{-1/2} \bar \xi^j_{-1/2}|0\rangle_{\textsc{ns}}$ from the free-fermionic sector.\footnote{In the $(4,2)$ coset discussed 
above, one has in particular the R-current of the right moving $\mathcal{N}=2$ algebra.} As this string state  
has an $\slr$ spin $J=0$ it is not normalizable. It shows that in the double scaling limit the wave-functions 
of the unbroken gauge group bosons do not have support near the bolt of Eguchi--Hanson. They correspond to a {\it global} symmetry 
of the interacting theory localized on the two-cycle.

\subsection{Worldsheet non-perturbative effects} 
\label{nonpert}
In supergravity, the  Eguchi--Hanson instanton is identified with the resolution 
of an $A_1$ singularity; the $\mathbb{Z}_2$ orbifold is necessary in order to avoid 
a conical singularity at $r=a$. So far we did not find such a constraint from the worldsheet construction. 
By analogy with the gravity analysis, it should involve physical effects localized in target space near 
the bolt of the manifold. 

Considering first the $(4,1)$ model, one of the building blocks of the heterotic string background~(\ref{cftback}) is the 
super-coset $\slc$. It is known~\cite{fzz,Kazakov:2000pm,Hori:2001ax} that this superconformal field theory receives 
worldsheet non-perturbative corrections in the form of an $\mathcal{N}=(2,2)$ Liouville potential. 

Asymptotically one can view the worldsheet CFT~(\ref{cftback}) as the  $\mathbb{R}_Q \times SU(2)_k/\mathbb{Z}_2$ theory 
perturbed by the sigma-model deformation:
\begin{equation}
\delta S 
=\mu \int \di^2 z \, \mathrm{e}^{-\frac{\varrho}{\ell}} \big(J^3  + 
:\psi^\varrho \psi^3:\big) (\ell :\bar \xi^3 \bar \xi^4: +:\bar \xi^1 \bar \xi^2:) \, ,
\label{cigint}
\end{equation}
that follows our general ansatz~(\ref{dyndef}), with extra fermionic interactions requested by 
worldsheet supersymmetry. This perturbation corresponds in the analysis done in~\ref{massless} to the 
$(a,a)\otimes \bar{a}_u$ operator in the singlet of G. In addition the worldsheet action of this \textsc{cft} 
is corrected quantum mechanically with an asymmetric $\mathcal{N}=(2,2)$ Liouville  potential.
In order to write this marginal interaction term in the present context, 
we parametrize the $J^3$ and $\bar{\jmath}^g$ currents in terms of chiral bosons as 
\begin{equation}
J^3 = i\sqrt{\tfrac{k}{\alpha'}}\, \partial Y_\textsc{l} \quad , \qquad 
\bar{\jmath}^g = i:\bar \xi^3 \bar \xi^4: = i \sqrt{\tfrac{2}{\alpha'}}\, \bar{\partial} X_\textsc{r}.
\end{equation}
Then the dynamically generated Liouville potential reads
\begin{equation}
\delta S = \mu_{\mathrm{L}} \int \di^2 z \, (\psi^\varrho + i\psi^3)(\bar{\xi}^1-i\bar{\xi}^2) \, 
\mathrm{e}^{-\ell (\varrho +i [Y_\textsc{l}+X_\textsc{r}])}+ \textrm{c.c.} \, , 
\label{liouvint}
\end{equation}
corresponding to the singlet twisted state $(c,c)\otimes \bar{c}_t$. 
The $SU(2)/U(1)$ contribution on the left-moving side is actually trivial, as can be seen using 
the character identity~(\ref{reflsym}). It also preserves the full $\widehat{\mathfrak{su}(2)}_{k-2}$ 
symmetry of the right-movers, corresponding to the isometries of the two-sphere in Eguchi--Hanson.

The very existence of the CFT at  non-perturbative level requires this Liouville operator 
to be in the physical spectrum, at left-moving superghost number zero.\footnote{
This is indeed the correct super-ghost picture needed to write a perturbation of sigma-model action, integrated over 
the $(1,0)$ string super-worldsheet.} We have found in subsec.~\ref{massless} that it has to belong to the {\it twisted sector} 
of the $SU(2)_k/\mathbb{Z}_2$ orbifold. In other words, the orbifold is necessary for the non-perturbative
consistency of the worldsheet theory; this mirrors the condition of no conical singularity in supergravity. 
Furthermore, the existence of such a state in the twisted sector dictates the condition\footnote{
In this case the sign factor $(-)^{\delta\gamma (k/2-1)}$ becomes trivial in the partition function~(\ref{partorb}). It corresponds to the distinction between $D_{2n+2}$ and $D_{2n+1}$ invariants of 
$SU(2)_k$.} $k-2\equiv 0 \mod 4$.
The heterotic \textsc{gso} projection provides another constraint for the existence 
of the Liouville potential~(\ref{liouvint}), as its right \textsc{gso} parity is $(-)^{\ell+1}$. 
Both conditions are satisfied provided that $\ell$ is an odd-integer.

This coset \textsc{cft} has an enhanced $\mathcal{N}=(4,1)$ superconformal symmetry. 
The  left-moving part of the Liouville potential~(\ref{liouvint}) (that is actually 
identical to the holomorphic side of the symmetric model discussed in~\cite{Israel:2004ir}) 
preserves an extended  $\mathcal{N}=4$ superconformal symmetry, 
implying eight supercharges in space-time. On the right-moving side, this Liouville potential 
preserves an $\mathcal{N}=2$ superconformal symmetry.

\subsubsection*{Generic bundle}
One can carry on a similar analysis of worldsheet non-perturbative effects for a generic $U(1)^{16}$ bundle over Eguchi--Hanson with 
a shift vector $\vec{\ell}$. One 
first has to bosonize the Cartan gauge currents fermion bilinears:
\begin{equation}
 : \!\bar \xi^{2i+1} \bar \xi^{2i+2}\!:\, = \sqrt{\tfrac{2}{\alpha'}}\,\bar \partial X^i_\textsc{r} \quad i=0,\ldots,15.
\end{equation} 
In a similar way this  coset \textsc{cft}  receives non-perturbative corrections, in 
the form of an $\mathcal{N}=(2,0)$ Liouville interaction, whose right-moving part is 
actually similar to a Sine-Liouville potential. It belongs also to the twisted sector of the $\mathbb{Z}_2$ orbifold. 
Using the quantization condition~(\ref{quantbis}) one obtains the interaction term:
\begin{equation}
\delta S = \mu_{\mathrm{L}} \int \di^2 z \, (\psi^\varrho + i\psi^3) \, 
\mathrm{e}^{-\sqrt{\vec{\ell}^2-1} (\varrho +i Y_{\mathrm{L}})-i \vec{\ell}\cdot \overrightarrow{X}_\textsc{r}}+ \mathrm{c.c.} \, , 
\label{multliouvint}
\end{equation}
which is indeed marginal. This operator is part of the spectrum only if $k-2\equiv 0 \mod 4$ as discussed above. 
The right-moving \textsc{gso} parity of this operator is given by $(-)^{\sum_i \ell_i}$. Hence to ensure that 
the Liouville operator corresponding to~(\ref{multliouvint}) belongs to the physical spectrum 
one has to satisfy
\begin{equation}
\sum_{i} \ell_i \equiv 0 \mod 2 \, ,
\end{equation}
which solves simultaneously these two constraints. 

Remarkably, this corresponds exactly to the so-called K-theory constraint 
for the stability of the  gauge bundle $V$~\cite{Witten:1985mj,Freed:1986zx,Blumenhagen:2005ga}, which applies to the 
first Chern class: 
\begin{equation}
c_1 (V) \in H^2 (\textsc{eh},2\mathbb{Z}) \implies \sum_{i} \int_{\Sigma} \frac{\mathcal{F}^i}{2\pi} =  \sum_{i} \ell_i \equiv 0 \mod 2 \, ,
\label{ktheory}
\end{equation}
integrating  over the two-cycle $\Sigma$ of Eguchi--Hanson. Under this condition the gauge bundle admits spinors that arise in 
the massive spectrum of the heterotic string. It can be seen directly in our construction, looking at eq.~(\ref{ferbundle}), 
that the spinorial representations of $Spin(32)/\mathbb{Z}_2$ are orbifold-invariant provided~(\ref{ktheory}) holds.

\subsection{Models with no global tadpoles}
We will now apply the methods developed in this work to the particular examples of purely Abelian 
gauge bundle over Eguchi--Hanson where the tadpole condition is fulfilled, i.e. such that the Bianchi identity is 
satisfied globally, in the asymptotically flat case, but not locally. Bundles  of this sort were considered in the supergravity 
approach in~\cite{Nibbelink:2007rd}; the massless spectrum was derived from the anomaly polynomial using 
the methods of~\cite{Honecker:2006qz}.

Taking the double scaling limit, one obtains a subspace of  our models that satisfies $\vec{\ell}^2 = 6$. Interestingly, 
even though the tadpole condition is satisfied, one still finds a solitonic fivebrane-like object 
on the neighborhood of the two-cycle (the Bianchi identity 
cannot be satisfied locally for a purely Abelian bundle as explained in sec.~\ref{secsugra}).  
Setting aside the spinorial shift case (that can be considered as well using these methods) there are two possible Abelian bundles, 
see table~1 in reference~\cite{Nibbelink:2007rd}.

In the first example the coset \textsc{cft} is constructed using the affine current 
\begin{equation}
\bar{\jmath}_g =  \vec{\ell} \cdot \bar{\partial} \overrightarrow{X}_\textsc{r}
\end{equation}
with shift vector
\begin{equation}
\vec{\ell}=\Big(\underbrace{1,\ldots,1}_{\times 6},\underbrace{0,\ldots,0}_{\times 10}\Big) \, .
\end{equation}
The affine currents commuting with $\bar{\jmath}_g$  define an $SO(20)_1 \times SU(6)_1$ affine algebra corresponding to 
the unbroken gauge symmetry. As explained above in the double scaling limit this gauge symmetry is merely 
a global symmetry of the interacting degrees of freedom.  The $U(1)$ gauge boson associated to $\bar{\jmath}_g$ has 
a mass $\mathfrak{m} =2/\sqrt{5\alpha'}$, close to the string scale.  

To write the explicit form of the operators, it is convenient to introduce the 16-dimensional vector $\vec{n}$ 
whose components $n_i$ denote the charge associated with the current $: \!\bar \xi^{2i+1} \bar \xi^{2i+2}\!:$ 
from the Cartan of the affine $SO(32)_1$ algebra. In our particular example the components with $i>6$ are set to zero. 
The $\slr$ quantum number $M$ for a given state is given by $M=\vec{\ell} \cdot \vec{n}$.  In order to classify the 
different massless localized states that appear in the spectrum we write the $\slc$ part of the vertex operators in a 
free-field representation, valid in the asymptotic region $\varrho \to \infty$. For unitary normalizable 
discrete representations, the $\slr$ spin $J$ lies in the range $\nicefrac{1}{2} < J < \nicefrac{11}{2}$. 
The $SU(2)_{\mathrm{L}}\times SU(2)_{\mathrm{R}}$ primaries are denoted $V_{j m \bar m}$ and $\mathrm{e}^{-\varphi}$ is the left-moving 
super-ghost contribution in the \textsc{ns} sector. 
\\

\noindent $\diamond$
The first type of localized operator one considers  is of the asymptotic form (for large $\varrho$)
\begin{equation}\label{V1}
\mathcal{V}_1 =  \mathrm{e}^{-\varphi} \mathrm{e}^{ip_\mu X^\mu} \, 
\mathrm{e}^{-\frac{J \, \varrho}{\sqrt{\vec{\ell}^2-1}}-i \vec{n} \cdot \overrightarrow{X}_\textsc{r} }\,
 \,V_{J-1;M,\bar m}\, .
\end{equation}
This coset operator comes from a primary state of $\slr$ in the discrete representations, hence the charge
$M$ and the spin $J$ are related as $M=J+r$ with $r\in \mathbb{N}$. In the untwisted sector, the mass-shell
condition imposes $\vec{n}^2=2$. The solutions are $(J,M) \in \{(1,2);(2,2)\}$. These two hypers
$(\mathbf{1},\mathbf{15})$ are singlets of $SO(20)$ and in the antisymmetric representation of $SU(6)$. 
In the twisted sector of the $\mathbb{Z}_2$ orbifold, the mass-shell condition reads $\vec{\ell}^2-2J + \vec{n}^2 = 2$. 
The first solution is $\vec{n}^2=6$ with $(J,M)=(5,6)$ the second one $\vec{n}^2=4$ with $(J,M)=(4,4)$. One obtains
then the hypermultiplets $(\mathbf{1},\mathbf{1}) + (\mathbf{1},\mathbf{15})$. The former is nothing 
but the Liouville operator~(\ref{multliouvint}).\\

\noindent $\diamond$
The second type of operator  is of the form
\begin{equation}\label{V2}
\mathcal{V}_2 =  \mathrm{e}^{-\varphi} \mathrm{e}^{ip_\mu X^\mu}\,  \bar{\xi}^\alpha \,\mathrm{e}^{-\frac{J\,\varrho}{\sqrt{\vec{\ell}^2-1}}-i \vec{n} \cdot \overrightarrow{X}_\textsc{r}}\, V_{J-1;M,\bar m}\,,
\end{equation}
with $\alpha>6$, i.e. tensoring a $\bar \Delta = \nicefrac{1}{2}$ primary of the coset (coming also from an 
$\slr$ primary) with a fermionic oscillator from the unbroken $SO(20)_1$ algebra. In the untwisted sector, 
the mass-shell condition $\vec{n}^2=1$ has the solution $(J,M)=(1,1)$ giving a bi-fundamental state 
$(\mathbf{20},\mathbf{6})$. In the twisted sector, the mass-shell condition $\vec{\ell}^2-2J + \vec{n}^2 = 1$ 
has the solution $(J,M) = (5,5)$ giving a hypermultiplet $(\mathbf{20},\mathbf{6})$.\\

\noindent $\diamond$
The last type of localized operator  is of the asymptotic form
\begin{equation}\label{V3}
\mathcal{V}_3 =  \mathrm{e}^{-\varphi} \mathrm{e}^{ip_\mu X^\mu} \,  \vec{\ell} \cdot \db \vec{X}_\textsc{r} \,\mathrm{e}^{-\frac{J \, \varrho}{\sqrt{\vec{\ell}^2-1}}-i \vec{n} \cdot \overrightarrow{X}_\textsc{r} }\,V_{J-1;M,\bar m}\, .
\end{equation}
This coset operator comes from a descendant state of $\slr$ of the form $K^{-}_{-1} |J,M\rangle$, hence 
it satisfies $M=J+r-1$ with $r\in \mathbb{N}$. In the untwisted sector, the mass-shell 
condition imposes $\vec{n}^2=0$, so $(J,M)=(1,0)$. This singlet hypermultiplet $(\mathbf{1},\mathbf{1})$ corresponds to the
dynamical current-current deformation, generalizing~(\ref{cigint}). In the twisted sector, the mass-shell condition  now reads
$\vec{\ell}^2-2J + \vec{n}^2 = 0$. The only solution is $\vec{n}^2=4$ with $(J,M)=(5,4)$, giving the hypermultiplet 
$(\mathbf{1},\mathbf{15})$. \\

\begin{figure}[!!t]
\centering
\begin{tabular}{|c||c|c||l|}\hline
$\vec{\ell}$ & Untwisted sector & Twisted sector & Gauge bosons\\
\hline
$(1^6,0^{10})$&$2(\mathbf{1},\mathbf{15})$ & $2(\mathbf{1},\mathbf{15})$ & Massive $U(1)$, mass $\mathfrak{m}=\tfrac{2}{\sqrt{5\alpha'}}\,.$\\
&$(\mathbf{20},\mathbf{6})$ & $(\mathbf{20},\mathbf{6})$ & $SO(20)\times SU(6)\,,$\\
&$(\mathbf{1},\mathbf{1})$ & $(\mathbf{1},\mathbf{1})$ & non-normalizable \\[2pt]
\hhline{=::==::=}
$(1^2,2,0^{13})$& $2(\mathbf{1},\mathbf{1})_2$ & $2(\mathbf{1},\mathbf{1})_{-2}$ & Massive $U(1)$, mass $\mathfrak{m}=\tfrac{2}{\sqrt{5\alpha'}}\,.$\\
&$3(\mathbf{1},\mathbf{2})_0$ &    $2(\mathbf{1},\mathbf{2})_{0}+(\mathbf{1},\mathbf{2})_2$ & $\ $ \\
&$2(\mathbf{26},\mathbf{1})_{-1}$ & $(\mathbf{26},\mathbf{1})_{1}$& $SO(26)\times SU(2)\times U(1)\,,$\\
&$(\mathbf{26},\mathbf{2})_{1}$ & $(\mathbf{26},\mathbf{2})_{-1}$ & non-normalizable\\
& $(\mathbf{1},\mathbf{1})_0$ & $(\mathbf{1},\mathbf{1})_0$ &
\\[2pt]
\hline
\end{tabular}
\caption{\it Spectra of hypermultiplets and gauge bosons for the two tadpole-free line bundles with vector structure.}
\label{tadfree}
\end{figure}

The second example with integer shift listed in~\cite{Nibbelink:2007rd} corresponds to a model with
shift vector
\begin{equation}
\vec{\ell}=\Big(1,1,2,\underbrace{0,\ldots,0}_{\times 13}\Big) \, ,
\end{equation}
leading to an unbroken $SO(26)\times SU(2)\times U(1)$ gauge symmetry. The charges under 
the massless $U(1)$ factor are determined by $\widetilde{Q}=\vec{n}\cdot \vec{w}$, where $\vec{w}=(1^2,-1,0^{13})$ is orthogonal to the shift vector $\vec{\ell}$. This accounts in particular for
the fact that the two universal hypermultiplets corresponding to the Liouville operator~(\ref{multliouvint}) and the dynamical current-current deformation~(\ref{cigint}) are uncharged, as displayed in the last line of tab.~\ref{tadfree}. The rest of the hypermultiplet spectrum can be determined by repeating the analysis performed in the previous example.\\

\noindent $\diamond$
For the operators of type $\mathcal{V}_1$ (\ref{V1}), the untwisted sector yields $2(\mathbf{1},\mathbf{1})_{2}+3(\mathbf{1},\mathbf{2})_{0}$, resulting respectively from $M=2$ and $M=3$ primaries of $\slr$. The twisted sector, on the other hand, leads to $(\mathbf{1},\mathbf{2})_{0}+(\mathbf{1},\mathbf{1})_{-2}+
(\mathbf{1},\mathbf{2})_{2}+(\mathbf{1},\mathbf{1})_{0}$, resulting from $\slr$ primaries of charge
$M=3,4,5,6$.\\

\noindent $\diamond$
For the operators $\mathcal{V}_2$ (\ref{V2}), we obtain the hypermultiplets $(\mathbf{26},\mathbf{2})_{1}+2(\mathbf{26},\mathbf{1})_{-1}$ in the untwisted sector, with $\slr$ charges $M=1,2$, while the twisted sector yields $(\mathbf{26},\mathbf{1})_{1}+(\mathbf{26},\mathbf{2})_{-1}$, with charges $M=4,5$.\\

\noindent $\diamond$
Finally, the operators $\mathcal{V}_3$ (\ref{V3}) contribute to the hypermultiplet spectrum as $(\mathbf{1},\mathbf{1})_{0}$ in the untwisted sector, with $M=0$, and $(\mathbf{1},\mathbf{2})_{0}+(\mathbf{1},\mathbf{1})_{-2}$ in the twisted sector, with respectively $M=3,4$.\\

We summarize the full spectra of these two examples in tab.~\ref{tadfree}. In the $\vec{\ell}=(1^6,0^{10})$ model, we observe an (accidental) symmetry of representations between the untwisted and
twisted sector of the orbifold.\footnote{This is clearly different from what is seen, for instance, 
in the perturbative $\mathbb{C}^2/\mathbb{Z}_2$ orbifold.} 

The singlet hypermultiplets in the third and eighth line are especially important as they correspond to the current-current
operator and the Liouville operator respectively, that acquire simultaneously a non-zero vacuum expectation 
value in the blow-up regime. This pair of hypermultiplets appears in the 
spectrum for all Abelian gauge bundle of the type we considered in this work. They correspond respectively to the modulus for the volume of the two-cycle 
and to the blow-up mode.

Comparing with the supergravity results of~\cite{Nibbelink:2007rd} we observe that
the multiplicities of hypermultiplets differ in all cases except for the states in the bi-fundamental (however, the gauge group representations appearing in the spectrum are the same).
It is certainly related to the fact that the blow-down limit is different in our models and those of reference~\cite{Nibbelink:2007rd}. In 
the latter case the blow-down limit was taken as the non-singular orbifold, i.e. the spectrum was continuously connected to the spectrum of heterotic strings at the standard orbifold point $\mathbb{C}^2/\mathbb{Z}_2$. 
In our case one gets the genuine $A_1$ singularity, with a non-trivial interacting Little String Theory emerging rather than an ordinary orbifold worldsheet 
\textsc{cft}. Clearly, the Abelian gauge bundle data, which is the same in both cases, is not sufficient to fully characterize 
the models. This comparison clearly deserves further study.\footnote{We thank M.~Trapletti for discussions on these issues.}

\subsection*{Acknowledgments}

The authors would like to thank C.~Bachas, E.~Dudas, S.~Groot-Nibbelink, C.~Kounnas, V.~Niarchos, M.~Petrini,
N.~Prezas, K.~Sfetsos, J.~Sonnenschein and especially M.~Trapletti
for numerous scientific discussions. They also acknowledge N.
Prezas' contribution at the first stages of the project. 

This work was supported in part by the EU under the contracts
MEXT-CT-2003-509661, MRTN-CT-2004-005104, MRTN-CT-2004-503369,
by the Agence Nationale pour la Recherche, France, under contract
05-BLAN-0079-01, by the Conseil r\'egional d'Ile de France, under convention N$^{\circ}$F-08-1196/R , by the Swiss National Science Foundation, under contract PBNE2-110332,
and by the Holcim Stiftung zur F\"orderung der wissenschaftlichen Fortbildung.

Marios Petropoulos would like to thank the Institut de Physique de
l'Universit\'e de Neuch\^atel where this work was initiated and
acknowledges partial financial support by the Swiss National
Science Foundation.

\appendix

\section{Heterotic supergravity: connections, curvatures and equations of motion}\label{appsigma}

We give here the calculations of the relevant geometrical quantities for sec. \ref{secsugra}.
\boldmath
\subsection*{The Eguchi--Hanson space}
\unboldmath
\noindent
In the conventions for the $SU(2)$ left-invariant one-forms given in expr.~(\ref{su2-1forms}),  the vierbein associated to the Eguchi--Hanson metric~(\ref{EHmetric}) reads
\begin{equation}\label{vier}
\hat{e}^a=\left\{
\frac{\di r}{g(r)},\frac{r\,\sigma_1^{\textsc{l}} }{2},\frac{r\,\sigma_2^{\textsc{l}} }{2}, \frac{rg(r)\,\sigma_3^{\textsc{l}} }{2}
\right\}\,.
\end{equation}
where the frame indices are denoted by $a,b={0,1,2,3}$ and $i,j={1,2,3}$, and the function
$$
g(r)^2 = 1-\left(\frac{a}{r}\right)^4
$$
is responsible for the deformation of the $S^1$ fiber of the \textsc{eh} space, as one navigates along the radial coordinate.
The $\sigma_i^{\textsc{l}}$ are the $\su$ left-invariant one-forms defined on $S^3$ and given explicitely in (\ref{su2-1forms}). They satisfy the Maurer-Cartan equation:
$$
\di \sigma_i^{\textsc{l}} = \varepsilon_{i\phantom{jk}}^{\phantom{i}jk}\,\sigma_j^{\textsc{l}}\wedge \sigma_k^{\textsc{l}}\,.
$$

\noindent
The  orientation of the volume measure on \textsc{eh} is chosen in accordance with $\varepsilon_{r\psi_{\mathrm{L}}\theta\psi_{\mathrm{R}}}=+1$:
\begin{equation}\label{EHvol}
\Omega_{\textsc{eh}}= \hat{e}^0\wedge\hat{e}^1\wedge\hat{e}^2\wedge\hat{e}^3 =
 \left(\frac{r}{2}\right)^3\,\sin\theta\, \di r \wedge \di\psi_{\mathrm{L}} \wedge \di\theta\wedge \di\psi_{\mathrm{R}}\,.
\end{equation}

\noindent
The spin connection for (\ref{vier}) reads:
\begin{equation}
\label{omegaEH}
\begin{array}{lll}
{\ds \hat{\omega}^{0}_{\phantom{0}1}= -\frac{g}{r}\, \hat{e}^1\,,} &
{\ds \hat{\omega}^{0}_{\phantom{0}2}= -\frac{g}{r}\, \hat{e}^2\,,} & \qquad
{\ds \hat{\omega}^{0}_{\phantom{0}3}= \frac{g^2-2}{r g}\, \hat{e}^3\,,} \\[6pt]
{\ds \hat{\omega}^{i}_{\phantom{0}j}= -\varepsilon^{i\phantom{j}k}_{\phantom{i}j}\, \hat{\omega}^{0}_{\phantom{0}k} }\, & &
\end{array}
\end{equation}
and is anti-selfdual in the frame indices.

\noindent
The curvature constructed from (\ref{omegaEH}) is given by:
\begin{equation}
\label{R2EH}
\begin{array}{ll}
{\ds \mathcal{R}^{0}_{\phantom{0}1}=-\frac{2(1-g^2)}{r^2}\,\big( \hat{e}^0\wedge \hat{e}^1- \hat{e}^2\wedge \hat{e}^3\big)}\,, &\qquad
{\ds \mathcal{R}^{0}_{\phantom{0}2}=-\frac{2(1-g^2)}{r^2}\,\big( \hat{e}^0\wedge \hat{e}^2+ \hat{e}^1\wedge \hat{e}^3\big)}\,, \\[6pt]
{\ds \mathcal{R}^{0}_{\phantom{0}3}=\frac{4(1-g^2)}{r^2}\,\big( \hat{e}^0\wedge \hat{e}^3- \hat{e}^1\wedge \hat{e}^2\big)}\,, &\\[6pt]
{\ds \mathcal{R}^{i}_{\phantom{0}j}=-\varepsilon^{i\phantom{j}k}_{\phantom{i}j} \, \mathcal{R}^{0}_{\phantom{0}k} }\,. & 
\end{array}
\end{equation}
It is anti-selfdual both in the frame and the coordinate indices.

\boldmath
\subsection*{Heterotic fivebranes on Eguchi--Hanson space}
\unboldmath
\noindent
We write the conformal factor for fivebranes transverse to \textsc{eh} space in the general form:
$$
H(r)= \lambda+\frac{2\alpha'Q_5}{r^2}\,,
$$
in terms of a two-valued parameter parameter $\lambda=0,1$, giving respectively the near horizon limit and the asymptotically flat solution of the heterotic fivebrane.

\noindent
In this case, the generalised spin connections with torsion reads:
\begin{equation}
\label{omega-}
\begin{array}{ll}
{\ds \Omega^{\phantom{+}0}_{-\phantom{0}1}= -\frac{\lambda g}{rH}\,\hat{e}^1\,,} & \qquad
{\ds \Omega^{\phantom{+}0}_{-\phantom{0}2}= -\frac{\lambda g}{rH}\,\hat{e}^2\,,} \\[12pt]
{\ds\Omega^{\phantom{+}0}_{-\phantom{0}3}= -\frac{2H(1-g^2)+\lambda g^2}{rgH}\,\hat{e}^3\,,} &\qquad
{\ds\Omega^{\phantom{+}1}_{-\phantom{0}2}= \frac{2H-\lambda g^2}{rgH}\,\hat{e}^3\,,} \\[12pt]
{\ds\Omega^{\phantom{+}1}_{-\phantom{0}3}= g\left(\frac{\lambda-2H}{rH}\right)\,\hat{e}^2\,,} &
{\ds\Omega^{\phantom{+}2}_{-\phantom{0}3}= g\left(\frac{2H-\lambda}{rH}\right)\,\hat{e}^1\,,}
\end{array}
\end{equation}
which has no particular duality symmetry in the frame indices, while:
\begin{equation}
\label{omega+}
\Omega^{\phantom{+}0}_{+\phantom{0}1}= \Omega^{\phantom{+}0}_{-\phantom{0}1}\,,\qquad
\Omega^{\phantom{+}0}_{+\phantom{0}2}= \Omega^{\phantom{+}0}_{-\phantom{0}2}\,,\qquad
\Omega^{\phantom{+}0}_{+\phantom{0}3}= \Omega^{\phantom{+}0}_{-\phantom{0}3}\,,\qquad
\Omega^{\phantom{+}a}_{+\phantom{0}b}= -\varepsilon^{a\phantom{b}c}_{\phantom{a}b}\,\Omega^{\phantom{+}0}_{+\phantom{0}c}\,
\end{equation}
is anti-selfdual in the latter.

\noindent
As we have seen in sec.\ref{secsugra}, only $\Omega_-$ appears in the Bianchi identity (\ref{bianchi}). We give here the relevant curvature two-form:
\begin{subequations}
\begin{align} \notag
\mathcal{R}^{\phantom{+}0}_{-\phantom{0}1}&= 
\frac{2}{r^2H^2}\Big[ \lambda\big[\lambda g^2-H\big]\,\hat{e}^0\wedge\hat{e}^1 + \big[H(2H-\lambda)(1-g^2)+\lambda(H-\lambda)g^2\big]\,\hat{e}^2\wedge\hat{e}^3 \Big]
\,,\\ \notag
\mathcal{R}^{\phantom{+}0}_{-\phantom{0}2}&= 
\frac{2}{r^2H^2}\Big[ \lambda\big[(\lambda g^2-H\big]\,\hat{e}^0\wedge\hat{e}^2 - 
\big[H(2H-\lambda)(1-g^2)+\lambda(H-\lambda)g^2\big]
\,\hat{e}^1\wedge\hat{e}^3 \Big]
\, ,\\ \notag
\mathcal{R}^{\phantom{+}0}_{-\phantom{0}3}&= 
\frac{2}{r^2H^2}\Big[ \big[ \lambda(\lambda g^2-H)+(1-g^2)H(4H-\lambda)\big]\,\hat{e}^0\wedge\hat{e}^3 
\\& \quad \qquad \qquad \qquad \qquad \qquad \qquad \qquad 
-\big[2H^2(1-g^2)+\lambda(\lambda-H)g^2\big]\,\hat{e}^1\wedge\hat{e}^2 \Big] \notag
\,,\\ \notag
\mathcal{R}^{\phantom{+}1}_{-\phantom{0}2}&= 
\frac{2}{r^2H^2}\Big[ \lambda\big[(H+\lambda)g^2-2H\big]\,\hat{e}^0\wedge\hat{e}^3 
+\big[2H^2(1-g^2)+\lambda(H-\lambda)g^2\big]\,\hat{e}^1\wedge\hat{e}^2 \Big]
\,,\\ \notag
\mathcal{R}^{\phantom{+}1}_{-\phantom{0}3}&= 
-\frac{2}{r^2H^2}\Big[ \big[2H^2(1-g^2)-\lambda(H-\lambda g^2)\big]\,\hat{e}^0\wedge\hat{e}^2 +\lambda\big[H-g^2(2H-\lambda)\big]\,\hat{e}^1\wedge\hat{e}^3 \Big]
\,,\\ \notag
\mathcal{R}^{\phantom{+}2}_{-\phantom{0}3}&= 
\frac{2}{r^2H^2}\Big[ \big[2H^2(1-g^2)-\lambda(H-\lambda g^2)\big]\,\hat{e}^0\wedge\hat{e}^1 -\lambda\big[H-g^2(2H-\lambda)\big]\,\hat{e}^2\wedge\hat{e}^3 \Big]
\,,
\end{align}
\end{subequations}

\noindent
\paragraph{The near horizon limit}
\noindent
We give here a more detailed presentation of some properties of the generalized spin connection and curvature two-form in the double scaling limit (\ref{DSL}), to support some statments made in sec.\ref{secsugra}. In particular, $\Omega_+$ simplifies to:
\begin{equation}
\Omega^{\phantom{+}0}_{+\phantom{0}3}=-\Omega^{\phantom{+}1}_{+\phantom{0}2}=\frac{2(g(r)^2-1)}{rg(r)}\,\hat{e}^3\,,
\end{equation}
while all other components of $\Omega_+$ vanish. In the near horizon limit, $\Omega_+$ then becomes Abelian, with curvature two-form:
$$
\mathcal{R}^{\phantom{-}0}_{+\phantom{0}3}=-\mathcal{R}^{\phantom{+}1}_{+\phantom{0}2}= \frac{4(1-g(r)^2)}{r^2}\big(2\hat{e}^0\wedge\hat{e}^3 - \hat{e}^1\wedge\hat{e}^2 \big)\,.
$$
The latter is anti-selfdual in the frame indices, but has no duality property in the coordinate ones.

\noindent
In the blowdown regime $g\rightarrow 1$, we readily verify that $\Omega_+\rightarrow 0$. From (\ref{omega-}), on the other hand,
we observe that $\Omega^{\phantom{-}0}_{-\phantom{0}i}=0$, $\forall i=1,..,3$, while the only surviving components are $\Omega^{\phantom{-}i}_{-\phantom{0}j}=\varepsilon^i_{\phantom{i}jk}\, \hat{e}^k$, for $i,j,k=1,..,3$, giving a flat connection. In the blowdown limit, both curvature two-forms $\mathcal{R}(\Omega_{\pm})$ vanish, as expected.

\boldmath
\subsection*{Heterotic equations of motion}
\unboldmath
\noindent
The equations presented here are beta-function equations for the heterotic string at lowest order in $\alpha'$ and are relevant to the computation of the dynamical promotion of the sigma-model (\ref{def-action-het}).

\noindent
We assume a heterotic string background described by a metric $g_{\mu\nu}$,
a Neveu--Schwarz two form $\mathcal{B}_{[2]}=\mathcal{B}_{\mu\nu}\, \mathrm{d}x^\mu\wedge
\mathrm{d}x^\nu$, a dilaton
$\Phi$ and a gauge field $\mathcal{A}_{\mu}$ that we assume to be Abelian, with coupling $k_g$. 
Conformal invariance requires the background fields to satisfy the
following equations:
\begin{subequations}\label{HETeom}
\begin{align}
0&=\mathcal{R}_{\mu\nu}^{\vphantom{g}} - \frac{1}{4}
\mathcal{H}^{\hphantom{m}\rho\sigma}_{\mu}\, \mathcal{H}^{\vphantom{g}}_{\nu \rho
\sigma} + 2 \,\nabla_{\mu}\nabla_{\nu}\Phi
-\frac{\alpha ' k_g}{4}\,\mathcal{F}_{\mu}^{\phantom{m}\rho}\mathcal{F}_{\nu\rho}
\,,\label{het1} \\
0&= \nabla_{\rho} \mathcal{H}_{\hphantom{m}\mu\nu}^{\rho}
-2 \,\mathcal{H}_{\hphantom{m}\mu\nu}^{\rho}\, \nabla_{\rho}\Phi
 \,,\label{het2} \\
0&=
\nabla_{\rho}\mathcal{F}^{\rho\mu}-2\,(\nabla_{\rho}\Phi)\mathcal{F}^{\rho\mu}+\frac{1}{2}\mathcal{H}^{\mu\kappa\lambda}\mathcal{F}_{\kappa\lambda} \,, \label{het3}
\\
c&=d + 3 \alpha' \left(4\,\nabla_{\rho}\Phi\, \nabla^{\rho}\Phi -2\,
\triangle \Phi  -\frac{1}{6} |\mathcal{H}|^2 -\frac{\alpha ' k_g}{8}|\mathcal{F}|^2\right) \,,\label{het4}
\end{align}
\end{subequations}
with $|\mathcal{G}|^2=\mathcal{G}^{\mu_1..\mu_p}\mathcal{G}_{\mu_1..\mu_p}$ for a general $p$-form. 

\noindent
The Abelian gauge and $\mathcal{B}$ curvatures are defined as:
\begin{equation}
\mathcal{F}_{[2]} = \di \mathcal{A}_{[1]} \quad , \qquad \mathcal{H}_{[3]} = \di \mathcal{B}_{[2]} +\tfrac{\alpha ' k_g}{4} 
\mathcal{A}_{[1]} \wedge \di \mathcal{A}_{[1]}.
\end{equation}
We will systematically neglect higher-order
corrections. Although the string models we are dealing with in the
present paper are exact, their background fields are in general known as expansions.

\section{${\mathcal N}=2$ characters and useful identities}
\label{appchar}
\boldmath
\subsection*{$\mathcal{N} =2$ minimal models}
\unboldmath
The characters of the $\mathcal{N} =2$ minimal models, i.e.  the supersymmetric $SU(2)_k / U(1)$ gauged \textsc{wzw} model, are conveniently defined through the characters $C^{j\ (s)}_{m}$~\cite{Gepner:1987qi} of the $[SU(2)_{k-2} \times U(1)_2] / U(1)_k$ bosonic 
coset, obtained by splitting the Ramond and Neveu--Schwartz 
sectors according to the fermion number mod 2. These characters are determined implicitly through the
identity:
\begin{equation}
\chi^{j} (\tau,\nu)
\Theta_{s,2}(\tau,\nu-\nu') = \sum_{m \in \zi_{2k}} C^{j\ (s)}_{m} (\tau,\nu')  \Theta_{m,k} (\tau,\nu-\tfrac{2\nu'}{k}) \, ,
\end{equation}in terms of the theta functions of $\widehat{\mathfrak{su} (2)}$
at level $k$, defined as 
\begin{equation}
\Theta_{m,k} (\tau,\nu) = \sum_{n}
q^{k\left(n+\tfrac{m}{2k}\right)^2}
e^{2i\pi \nu k \left(n+\tfrac{m}{2k}\right)}  \qquad m \in \mathbb{Z}_{2k}\, ,
\end{equation}
and $\chi^j (\tau,\nu)$ the  characters of the $\widehat{\mathfrak{su}(2)}$ affine 
algebra  at level $k-2$. Highest-weight representations are labeled by  $(j,m,s)$, corresponding primaries of 
$SU(2)_{k-2}\times U(1)_k \times U(1)_2$. The following identifications apply:
\begin{equation}
(j,m,s) \sim (j,m+2k,s)\sim
 (j,m,s+4)\sim
 (k/2-j-1,m+k,s+2)
\end{equation}
as  the selection rule $2j+m+s =  0  \mod 2$. The spin $j$ is restricted to $0\leqslant j \leqslant \tfrac{k}{2}-1$.  
The conformal weights of the superconformal primary states are:
\begin{equation}
\begin{array}{cclccc}
\Delta &=& \frac{j(j+1)}{k} - \frac{n^2}{4k} + \frac{s^2}{8} \ & \text{for} & \ -2j \leqslant n-s \leqslant 2j \\
\Delta &=& \frac{j(j+1)}{k} - \frac{n^2}{4k} + \frac{s^2}{8} + \frac{n-s-2j}{2}
\ & \text{for} & \ 2j \leqslant n-s \leqslant 2k-2j-4 \\
\end{array}
\end{equation}
and their $R$-charge reads:
\begin{equation}
Q_R = \frac{s}{2}-\frac{m}{k} \mod 2 \,. 
\end{equation}
A {\it chiral} primary state is obtained for $m=2(j+1)$ and 
$s=2$ (thus odd fermion number). It has conformal dimension
\begin{equation}
\Delta= \frac{Q_R}{2} = \frac{1}{2} - \frac{j+1}{k}\, .
\end{equation}
An {\it anti-chiral} primary state is obtained for $m=2j$ and $s=0$ 
(thus even fermion number). Its conformal dimension reads:
\begin{equation}
\Delta= -\frac{Q_R}{2} = \frac{j}{k}\, .
\end{equation}
Finally we have the following modular S-matrix for the $\mathcal{N}=2$ minimal-model characters:
\begin{equation}
S^{jm s}_{j' m' s'} = \frac{1}{2k} \sin \pi
\frac{(1+2j)(1+2j')}{k} \ e^{i\pi \frac{mm'}{k}}\ e^{-i\pi ss'/2}.
\end{equation}
The usual Ramond and Neveu--Schwarz characters, that we use in the bulk of the paper, are  obtained as:
\begin{equation}
C^{j}_{m} \oao{a}{b} =  e^{\frac{i\pi ab}{2}} \left[ C^{j\, (a)}_{m} 
+(-)^b C^{j\, (a+2)}_{m} \right],
\end{equation}
where $a=0$ (resp. $a=1$) denote the \textsc{ns} (resp. \textsc{r}) sector, and characters 
with $b=1$ are twisted by $(-)^F$. They are related to $\widehat{\mathfrak{su}(2)}_k$ characters 
through:
\begin{equation}
\chi^j  \vartheta \oao{a}{b} = \sum_{m \in \zi_{2k}} C^j_m \oao{a}{b} \Theta_{m,k}\,.
\end{equation}
In terms of those one has the reflexion symmetry:
\begin{equation}
C^j_m \oao{a}{b} = (-)^b C^{\tfrac{k}{2}-j-1}_{m+k} \oao{a}{b}\, . 
\label{reflsym}
\end{equation} 

\subsection*{Supersymmetric $\slc$}
The characters of the $\slc$ super-coset
at level $k$ come in different categories corresponding to
irreducible unitary representations of  $SL(2,\mathbb{R})$.

\noindent The \emph{continuous representations} correspond to $J = 1/2 + iP$,
$P \in \mathbb{R}^+$. Their characters are denoted by
 ${\rm ch}_c (\tfrac{1}{2}+ip,M) \oao{a}{b}$, where the $U(1)_R$ charge of the primary is $Q=2M/k$. They read:
\begin{equation}
{\rm ch}_c (\tfrac{1}{2}+ip,M;\tau,\nu) \oao{a}{b} = \frac{1}{\eta^3 (\tau)} q^{\frac{p^2+M^2}{k}} \vartheta \oao{a}{b} (\tau,\nu)
e^{2i\pi\nu \frac{2M}{k}}\, .
\end{equation}
The \emph{discrete representations}, of characters $\mathrm{ch}_d (J,r) \oao{a}{b}$,
have a real $\slr$ spin in the range $1/2 < J < (k+1)/2$. Their  $U(1)_R$  charge reads  $Q_R=2(J+r+a/2)/k$,
$r\in \zi$.  Their  characters are given by 
\begin{equation}
{\rm ch}_d (J,r;\tau,\nu) \oao{a}{b} =  \frac{
  q^{\frac{-(J-1/2)^2+(J+r+a/2)^2}{k}}
e^{2i\pi\nu \frac{2J+2r+a}{k}}}{1+(-)^b \,
e^{2i\pi \nu} q^{1/2+r+a/2} } \frac{\vartheta \oao{a}{b} (\tau, \nu)}{\eta^3 (\tau)}.
\label{idchar}
\end{equation}
One gets a {\it chiral} primary for $r=0$, i.e. $M=J$, in the \textsc{ns} sector  (with even fermion number). Its conformal dimension reads
\begin{equation}
\Delta= \frac{Q_R}{2} = \frac{J}{k}\, . 
\end{equation}An {\it anti-chiral} primary is obtained for $r=-1$ (with odd fermion number). Its conformal dimension reads
\begin{equation}
\Delta= -\frac{Q_R}{2} =\frac{1}{2}-\frac{J-1}{k}\, . 
\end{equation} 
\emph{Extended characters} are defined for $k$ integer by summing
over $k$ units of spectral flow~\cite{Eguchi:2003ik}.\footnote{One can extend their definition to the case of rational $k$, which 
is not usefull here.} For instance, the extended continuous characters are:
\begin{multline}
{\rm Ch}_c (\tfrac{1}{2}+ip,M;\tau,\nu) \oao{a}{b}=   \sum_{w \in \zi} 
{\rm ch}_c (\tfrac{1}{2}+ip,M+kw;\tau,\nu) \oao{a}{b} \\
= \frac{q^{\frac{p^2}{k}}}{\eta^3 (\tau)} \vartheta \oao{a}{b} (\tau,\nu)
\Theta_{2M,k} (\tau,\tfrac{2\nu}{k})\, ,
\label{extcontchar}
\end{multline}
where discrete $\mathcal{N}=2$  R-charges are chosen: $2M \in \zi_{2k}$. 
These characters close among themselves under the action of the modular group. 
For instance, the S transformation gives:
\begin{equation}
{\rm Ch}_{c} (\tfrac{1}{2}+ip,M;-\tfrac{1}{\tau}) \oao{a}{b}
= \frac{1}{2k}\int_0^\infty \!\!\! \di p' \, \cos \frac{4\pi p p'}{k}\!\! \sum_{2M' \in \mathbb{Z}_{2k}} \!\!
e^{-\frac{4i\pi M M'}{k}}
{\rm Ch}_{c} (\tfrac{1}{2}+ip',M';\tau) \oao{b}{-a}\, .
\end{equation}
The same holds for discrete representations, whose modular transformations are more involved 
(see~\cite{Eguchi:2003ik,Israel:2004xj}).


\bibliographystyle{JHEP}
\bibliography{bibbundle}

\end{document}